\documentclass[usenatbib, twocolumn]{rasti}

\usepackage{newtxtext, newtxmath}
\usepackage[T1]{fontenc}
\usepackage[acronym]{glossaries-extra}
\usepackage{natbib}
\usepackage{tabularx}
\usepackage{tikz}
\usepackage{bm}

\setabbreviationstyle[acronym]{long-short}
\newacronym{gw}{GW}{gravitational wave}
\newacronym{mdc}{MDC}{mock data challenge}
\newacronym{ligo}{LIGO}{Laser Interferometer Gravitational Wave Observatory}
\newacronym{us}{US}{United States}
\newacronym{gr}{GR}{general relativity}
\newacronym{pe}{PE}{parameter estimation}
\newacronym{hom}{HOM}{higher order mode}
\newacronym{fap}{FAP}{false alarm probability}
\newacronym{far}{FAR}{false alarm rate}
\newacronym{psd}{PSD}{power spectral density}

\glsdisablehyper

\usetikzlibrary{shapes.geometric, shapes.symbols, arrows, calc, positioning}

\tikzstyle{function} = [
  rectangle,
  text centered,
  draw=black
]

\tikzstyle{io} = [
  trapezium,
  trapezium stretches=true,
  trapezium left angle=70,
  trapezium right angle=110,
  text centered,
  draw=black
]

\tikzstyle{decision} = [
  signal,
  signal from=north,
  signal to=south,
  signal pointer angle=160,
  draw=black,
]

\tikzstyle{arrow} = [
  thick,
  ->,
  >=stealth
]

\title{LensingFlow: An Automated Workflow for Gravitational Wave Lensing
Analyses}
\author[Wright et. al]{
  Mick Wright$^{1, 2}$\thanks{Email: m.j.wright@uu.nl}, 
  Justin Janquart$^{3, 4}$\thanks{Email: justin.janquart@uclouvain.be}, 
  Paolo Cremonese$^{5}$, 
  Juno C. L. Chan$^{6}$, 
  Alvin K. Y. Li$^{7, 8, 9}$,
  \newauthor
  Otto A. Hannuksela$^{8}$,
  Rico K. L. Lo$^{6}$,
  Jose M. Ezquiaga$^{6}$,
  Daniel Williams$^{10}$,
  Michael Williams$^{11}$,
  \newauthor
  Gregory Ashton$^{12}$,
  Rhiannon Udall$^{9}$,
  Anupreeta More$^{13, 14}$,
  Laura Uronen$^{8}$,
  Ankur Barsode$^{15}$,
  \newauthor
  Eungwang Seo$^{10}$,
  David Keitel$^{5}$,
  Srashti Goyal$^{16}$,
  Jef Heynen$^{3}$,
  Anna Liu$^{8}$,
  Prasia Pankunni$^{13, 17}$
  \\
  $^{1}$ Institute for Gravitational and Subatomic Physics (GRASP), Department
  of Physics, Utrecht Univeristy,\\ Princetonplein 1, 3584 CC Utrecht, The
  Netherlands \\
  $^{2}$ Nikhef---National Institute for Subatomic Physics, Science Park, 1098
  XG Amsterdam, The Netherlands \\
  $^{3}$ Centre for Cosmology, Particle Physics, and Phenomenology---CP3,
  Universit\'{e} Catholique de Louvain, \\ Chemin du Cyclotron, 2---Box
  L7.01.05, B-1348 Louvain-la-Nueve, Belgium\\
  $^{4}$ Royal Observatory of Belgium, Avenue Circulaire, 3, B-1180 Brussels,
  Belgium \\
  $^{5}$ Department de F{\'i}sica, Universitat de les Illes Balears,
  IAC3--IEEC, E-07122 Palma, Spain\\
  $^{6}$ Center of Gravity, Niels Bohr Institute, Blegdamsvej 17, 2100
  Copenhagen, Denmark\\
  $^{7}$ RESCEU, The University of Tokyo, Tokyo, 113-0033, Japan\\
  $^{8}$ Department of Physics, The Chinese University of Hong Kong, Shatin, Hong
  Kong\\
  $^{9}$ LIGO Laboratory, California Institute of Technology, Pasadena,
  California 91125, USA\\
  $^{10}$ SUPA, School of Physics \& Astronomy, University of Glasgow, G12 8QQ
  Glasgow, Scotland\\
  $^{11}$ Institute of Cosmology and Gravitation, University of Portsmouth, PO1
  3FX Portsmouth, United Kingdom\\
  $^{12}$ Department of Physics, Royal Holloway, University of London\\
  $^{13}$ Inter-University Centre for Astronomy and Astrophysics (IUCCA), Post
  Bag 4, Ganeshkhind, Pune 411 007, India\\
  $^{14}$ Kavli IPMU (WPI), UTIAS, The University of Tokyo, Kashiwa, Chiba
  277-8583, Japan\\
  $^{15}$ International Centre for Theoretical Sciences, Tata Institute of
  Fundamental Researh, Banglaore 560089, India\\
  $^{16}$ Max Planck Insitute for Gravitational Physics (Albert Einstein
  Insitute)\\
  $^{17}$ Government Victoria College, Palakkad 678001, Kerala, India
}

\date{\today}
\pubyear{2025}

\begin{document}

\maketitle

\begin{abstract}
  In this work, we present \textsc{LensingFlow}. This is an implementation of an
automated workflow to search for evidence of gravitational lensing in a large
series of \glsfmtlong{gw} events. This workflow conducts searches for evidence
in all generally considered lensing regimes. The implementation of this
workflow is built atop the \textsc{Asimov} automation framework and
\textsc{CBCFlow} metadata management software and the resulting product
therefore encompasses both the automated running and status checking of jobs in
the workflow as well as the automated production and storage of relevant
metadata from these jobs to allow for later reproduction. This workflow
encompasses a number of existing lensing pipelines and has been designed to
accommodate any additional future pipelines to provide both a current and
future basis on which to conduct large scale lensing analyses of gravitational
wave signal catalogues. The workflow also implements a prioritisation
management system for jobs submitted to the schedulers in common usage in
computing clusters ensuring both the completion of the workflow across the
entire catalogue of events as well as the priority completion of the most
significant candidates. As a first proof-of-concept demonstration, we deploy
\textsc{LensingFlow} on a~\glsfmtlong{mdc} comprising $10$ signals in which
signatures of each lensing regime are represented. \textsc{LensingFlow}
successfully ran and identified the candidates from this data through its
automated checks of results from consituent analyses. 

\end{abstract}

\section{Introduction}\label{sec:intro}
In the intervening years between the initial detection of~\glspl{gw} in
2015~\citep{gw150914_discovery} and the publication of the most recent
icatalogue of events, GWTC-3~\citep{gwtc-3}, the current ground-based detector
network, consisting of the two~\glsfmtshort{ligo} detectors in
the~\glsfmtshort{us}~\citep{ligo_detector} and the Virgo detector in
Italy~\citep{virgo_detector}, has detected $\sim90$~\gls{gw} signals. As
sensitivities improve, additional detectors---such as
KAGRA~\citep{kagra_detector}---join the network, and both current and future
observing runs are carried out---such as the currently ongoing fourth observing
run (O4) which at the time of writing has already reported more than $200$
candidate detections~\citep{gracedb_detections}---this number will increase at
an accelerating rate~\citep{detector_observing_prospects}. This acceleration in
detections is expected to improve the current scientific analyses conducted
using~\glspl{gw}; such as cosmological parameter
estimation~\citep{o3_cosmology}, inferring the astrophysical population of
merging binaries~\citep{o3_randp}, constraint and observations~\citep{o3_tgr}
of any potential deviations from~\gls{gr}, etc.

To ensure that analyses keep pace with the rate of detections as it rapidly
increases, it will be important to allow for significant automation of the
deployment of these analyses. Indeed, to maintain the tractability of
Bayesian~\gls{pe} of incoming~\gls{gw} signals, the automation framework
\textsc{Asimov} was developed to allow for the large scale deployment of such
analyses~\citep{asimov}. Similarly, the vast numbers of individual analyses
each produce an amount of metadata that allows for the analysis to be open and
reproducible. Scaling this requires additional effort to ensure that it is well
organised and easily traversable for outsiders. The desire and need to achieve
this has led to the development of \textsc{CBCFlow}~\citep{cbcflow} to
standardise and organise metadata from current and future \gls{gw} analyses.

One area of \gls{gw}-based research in which the increasing rate of detections
will have a particularly sharp impact on the required computational resources
is the search for gravitational lensing of \glspl{gw}. Gravitational lensing
occurs when a signal passes close by to a massive object on the path between
the source and the observer. The warping of space-time around massive objects
that is a prediction of \gls{gr} affects the propagation of that passing
signal. The exact effects vary based on the nature of the object serving as the
lens but effects include: the production of multiple copies of the signal,
phase shifts, modulations of the amplitude, beating patterns from interference
and diffractions, and other distortions~\citep{Ohanian:1974ys, Thorne:1982cv,
Deguchi:1986zz, Nakamura:1997sw, Takahashi:2003ix}. Currently searches for
signatures of gravitational lensing have been conducted for those events up to
the release of GWTC-3~\citep{Hannuksela:2019kle, LIGOScientific:2021izm,
LIGOScientific:2023bwz, Janquart:2023mvf, Janquart:2024ztv,
Chakraborty:2025maj} but have not found confident evidence for detection. 

The multiple \gls{pe}-based analyses on a single \gls{gw} signal searching for
the individual distortions resulting from lensing---such as those proposed
in~\cite{Wright:2021cbn, Liu:2023ikc, Janquart:2021nus} respectively---would
already represent an increase in the number of required analyses to fully study
for evidence of signatures of gravitational lensing on that signal on a
per-waveform-model basis. However, the first mentioned effect of gravitational
lensing---the production of multiple signals---quadratically increases the
number of analyses required to investigate the lensing hypothesis as now not
only does every individual signal require analyses but the combinations of
signals must be analysed---such as through the \gls{pe}-based approaches
proposed in~\cite{Lo:2021nae} and~\cite{Janquart:2021qov}. 

In addition to the significant number of such analyses, one must also consider
the computational cost of each individual analysis. Performing \gls{pe}-based
approaches on individual signals is already a taxing challenge, and this is
only increased when considering combinations. To mitigate this, there exist
already a number of approaches to perform a low latency filtering of multiplets
to mitigate the need for the \gls{pe}-based analyses on every candidate
multiplet---see e.g.~\cite{haris_posterior_overlap, goyal_lensid, phazap,
Goyal:2023lqf}, and~\cite{Barsode:2024zwv}. Such low latency approaches are
much less computationally expensive than the \gls{pe}-based approaches, but are
less accurate. They therefore increase the tractability of the searches, but
add to the number of analyses that need to be performed as well as introducing
the problem of inter-pipeline communication to the workflow for lensing
analyses.

To address these potential roadblocks to performing complete lensing analyses
as the number and rate of \gls{gw} detections increases, we here present
\textsc{LensingFlow}~\citep{lensingflow}---an implementation of an automated
workflow for lensing analyses. It is built atop the \textsc{Asimov} automation
framework---to allow the analyses to be deployed easily at scale---and
leverages the metadata organisation capabilities of \textsc{CBCFlow} to ensure
both that its own output metadata is open and reproducible as well as that
prior unlensed analyses may be ingested and used by the lensing analyses
efficiently, accurately, and in a standardised fashion. Finally, it implements
both filtering and prioritisation of high latency \gls{pe} analyses based on
the recommendations from the lower latency investigations which ensures both
the completion of the entire catalogue of events and multiplets as well as the
rapid investigation of the most promising candidates. \textsc{LensingFlow}
consequently provides a solid foundation for lensing analyses both of actual
detector data as well as large-scale simulation studies.

We provide a proof-of-concept demonstration of \textsc{LensingFlow} by applying
it to a \gls{mdc} consisting of $10$ \gls{gw} signals. This set contains at
least one representation of each of the lensing regimes for which searches are
implemented in addition to unlensed signals.

The rest of this work will be laid out as follows. Section~\ref{sec:lensing}
provides a brief overview of the theory of gravitational lensing.
Section~\ref{sec:workflow} describes the layout of the workflow.
Section~\ref{sec:deployment} provides a detailed description of the \gls{mdc}
and the results from this deployment. Finally, section~\ref{sec:conclusion}
will summarise the conclusions of the work and provide insight into future
plans to further improve the automation of lensing analyses.

\section{Gravitational Lensing}\label{sec:lensing}
For a detailed description of gravitational lensing theory, we would refer the
reader to literature such as \cite{Schneider:1992bmb}. We will here present a
summary limited to the most relevant topics for this work and define the
terminology that will be used through-out this work.

As we have noted in Section~\ref{sec:intro}, gravitational lensing is a
consequence of a signal passing by a massive object, meaning that any
astrophysical object may be the source of lensing. As might be expected from
such a wide array of possible sources, the resulting phenomenology differs
significantly across the scales of the objects. Whilst lensing may be
structured based on the mass scale of the object, we will here define our
structuring of lensing-related terminology based upon how the analyses are
implemented. We will first note that there are two broad regimes of
lensing---those being where the \textit{geometric optics approximation} is
valid, and the effects of lensing may be considered purely in terms of the
geometry of the system, and the case where a full \textit{wave optics}
treatment is required where effects such as diffraction are playing a
noticeable role in the signatures imparted. The domain of the validity is
dependent upon a combination of the wavelength of the detected signal and the
mass of the lensing object. 

In the searches for lensing examined here, we consider the scenario in which
the primary effect of lensing is the production of multiple copies of that
signal---which we will term \textit{images} throughout the rest of the work. In
the regime where the geometric optics approximation is valid, these images each
have largely identical frequency evolution but are differentiated by an
individual \textit{magnification} of the signal amplitude, a \textit{time
delay} with respect to when the original unmodified signal would have arrived,
and an overall \textit{phase shift}~\citep{Schneider:1992bmb}. This means that
the observed \textit{strain} for the $j^{\textrm{th}}$ lensed image,
$h_{j}^{L}(f)$ may be given in terms of the unmodified strain produced by the
source, $h(f)$, as~\citep{Nakamura:1999uwi}
\begin{equation}
  \label{eq:geometric-optics-image}
  h^{L}_{j} \left( f; \bm{\theta}, \mu_{j}, \Delta{t_{j}}, n_{j} \right) =
  \sqrt{\left| \mu_{j} \right|} \exp \left(
      - 2\pi{i}f\Delta{t_{j}} + i\pi{n_{j}}
    \right) h\left(f; \bm{\theta} \right),
\end{equation}
where: $\bm{\theta}$ represents the set of parameters of the source binary;
$\mu_{j}$ the magnification of the image which itself may be defined in terms
of the Jacobian between the lens and source plane co-ordinate systems;
$\Delta{t_{j}}$ the time delay for the image resulting from a combination of
the geometrical path difference as well as Shapiro delay~\citep{shapiro_delay};
and $n_{j}$ the so-termed \textit{Morse index} which may take one of three
discrete values---$0$, $0.5$, and $1$. These values of the index correspond to
classification of the image as type I, II, or III respectively.

Where the signals are temporally resolvable as individual images, the
magnification and time delay do not cause any changes to the signal morphology.
Similarly Type I signals which correspond to the case of no overall phase shift
and Type III signals which correspond to the case of a sign flip in phase shift
which is completely degenerate with a $\pi/2$ shift in the polarisation angle,
also experience no effects on the signal morphology. By contrast, however, Type
II images experience an overall de-phasing between the modes of the signal
which may lead to a detectable distortion in waveforms that have significant
\gls{hom} content~\citep{Dai:2016igl, Ezquiaga:2020gdt, Wang:2021kzt,
Janquart:2021nus, Vijaykumar:2022dlp}.

If for a given detector will one decreases the mass of the lensing object, they
will reach the regime in which the time delay between the images is less than
the duration of the signal itself. If the lensing mass is still sufficiently
high that for this detector the geometric optics approximation is still valid,
one will have the case of apparently overlapping signals which can lead to the
entire set of images being detectred as one event with apparent beating
patterns in the combined signal. Given the condition that the geometric optics
approximation is still valid, this combined signal can be described as the
summation of the $j$ images described in
Equation~\ref{eq:geometric-optics-image}.

If we are in the case where the wavelength of the \gls{gw} signal is comparable
to the Schwarzschild radius of the lensing object, then the geometric optics
approximation will break down and a full wave optics treatment is required to
describe the effects on the signal. In this case, the effects of lensing are
entirely frequency-dependent modulations of the amplitidue of an apparent
signal that may provide insight into the nature of the lensing object
itself~\citep{Takahashi:2003ix, Cao:2014oaa, Lai:2018rto, Christian:2018vsi,
  Dai:2018enj, Jung:2017flg, Diego:2019lcd, Diego:2019rzc, Pagano:2020rwj,
Cheung:2020okf, mishra2021, Yeung:2021roe, Meena:2022unp, Wright:2021cbn}. The
exact morphology of the strain is highly dependent on the mass density
distribution of the lensing object but a general form is given
by~\citep{Nakamura:1999uwi, Takahashi:2003ix}
\begin{equation}
  \label{eq:wave-optics-amplification}
  F\left(w, \bm{y}\right) = \frac{w}{2\pi{i}} \int \exp \left[
    -iwT(\bm{x}, \bm{y})
  \right] d^{2}x,
\end{equation}
where $w$ is a dimensionless representation of the frequencies of the \gls{gw},
and $\bm{x}$ and $\bm{y}$ are dimensionless forms of the image and source
positions respectively and the integration is over the image plane. The
function $T$ computes the dimensionless time delay for a given image position
which will vary on a per-lens-model basis. We note here, for completion, that
taking the geometric optics approximation is to assert that only the stationary
points of the $T$ function are contribution to the result of the above
integral, which will yield the summation of
Equation~\ref{eq:geometric-optics-image} over the $j$ images. These stationary
points correspond to the individual images discussed when the approximation is
valid and the image type corresponds to the kind of stationary point resulting
in that image. We finally note that $w$ is defined such that the geometric
optics approximation is valid when $w \gg 1$.

\section{Analysis Workflow}\label{sec:workflow}
We now turn to describe the workflow including some brief descriptions of the
constituent lensing pipelines---though we would refer the reader again to the
literature for more specific details on each pipeline. These references are
summarised in Table~\ref{tab:pipelines}. During the following discussions, we
will assume that prior analyses have been done that covers the initial
detection of the~\gls{gw} event and continues through to standard
unlensed~\gls{pe} such as that, that would be produced for a catalogue such as
GWTC-3 and that the resulting metadata from these events has been used to
construct a \textsc{CBCFlow} library. In this work, we have also restricted the
multiplet analyses to operate only on the pairs of events for
simplicity---though we note that this restriction is not technical in nature
for the automated workflow. Consequently the following discussions may
naturally be extended to any set of multiplet. Some work has been done to make
extened multiplet analysis available in the package, though further work would
be required to bring this to equivalent functionality to pair-wise analyses.

\begin{table}
  \centering

  \caption{Constituent pipelines integrated within \textsc{LensingFlow} at the
    time of writing alongside the paper detailing their methodology and
    implementation. We note that \textsc{LensingFlow} is constructed to easily
    allow additional pipelines to be integrated in the future based on the
    needs of the \gls{gw} lensing community. N.B. in the case of
    \textsc{Atlenstics} whilst the pipeline implements some of the calculations
    detailed in the associated methodology paper, it was developed purely for
    convenient implementation of already existing work in the framework.}
  \label{tab:pipelines}

  \begin{tabularx}{0.8\columnwidth}{cX}
    \hline \hline Pipeline & Methodology Paper \\
    \hline \multicolumn{2}{c}{Multiplet Analyses} \\
    \hline \textsc{LensID} & \cite{goyal_lensid} \\
    \textsc{Phazap} & \cite{phazap} \\
    \textsc{Posterior\_Overlap} & \cite{haris_posterior_overlap} \\
    \textsc{Atlenstics*} & \cite{more_statistics} \\
    \textsc{Golum} & \cite{Janquart:2021qov} \\
    \textsc{hanabi} & \cite{Lo:2021nae} \\
    \hline \multicolumn{2}{c}{Single Event Analyses} \\
    \hline \textsc{Golum} & \cite{Janquart:2021nus} \\
    \textsc{Gravelamps} & \cite{Wright:2021cbn} \newline \cite{Liu:2023ikc} \\
    \hline
  \end{tabularx}

\end{table}

\subsection{Identification of Lensed Signals}\label{subsec:lensed-analyses}

The most complete means of identifying candidate lensed signals or multiplets
is identifying the features from each candidate regime and performing dedicated
\gls{pe} implementing those features---jointly for the multiple signals in the
case of multiplet analyses. This allows the computation of a Bayes factor
between the lensed and unlensed hypotheses. To calibrate the expectations of
values from performing these analyses using each of the dedicated \gls{pe}
models, one could then perform a background analysis over a large number of
unlensed injections into similar noise features as has been observed for the
data. This would then allow the construction of a distribution of these Bayes
factors for the unlensed population from which a \gls{fap} may be computed to
give the statistical significance of the candidate Bayes factor.

For lensing analyses, analysis frameworks to do such dedicated \gls{pe} based
searches have been developed both in the context of joint analysis of multiple
signals leading to the \textsc{Golum} and \textsc{hanabi}
pipelines~\citep{Janquart:2021nus, Janquart:2021qov, Janquart:2023osz,
Lo:2021nae} and in the context of single event analysis in both
modelled~\citep{Wright:2021cbn} and unmodelled~\citep{Liu:2023ikc} forms, both
of which are implemented in the \textsc{Gravelamps}
pipeline~\citep{gravelamps}.

Each of these aforementioned analyses implements their respective regime of
lensing---and in the case of \textsc{Gravelamps} multiple physical models of
lens as well as a phenomenological model---and includes wrapping functionality
around the \gls{pe} framework \textsc{bilby}~\citep{bilby_paper}, as well as
the \textsc{bilby\_pipe}~\citep{bilby_pipe_paper}---and
\textsc{parallel\_bilby}~\citep{pbilby_paper} in the case of
\textsc{hanabi}---high-throughput computing interfaces, menaing that the
resulting evidence from these investigations may be easily compared with
standard unlensed hypothesis outputs for model selection. In the cases of the
\textsc{Gravelamps} analyses and the \textsc{Golum} Type II analyses which
operate on a single signal basis, this comparison is already considered a final
Bayes factor for the hypothesis. In the cases of those analyses that operate on
multiple signals, one may additionally fold in population information and
estimates of selection effects to further refine the comparison to achieve a
final comparative Bayes factor~\citep{Lo:2021nae} under these specific
population and model assumptions. Consequently, following the lead of
e.g.~\cite{LIGOScientific:2023bwz} we will term the evidence comparison from
these joint \gls{pe} analyses, the \textit{coherence ratio}. 

In prior observing runs of the current detector network searches of these
regimes have been carried out~\citep{Hannuksela:2019kle,
LIGOScientific:2021izm, LIGOScientific:2023bwz} so far with no confident
detections. Performing these prior investigations has been possible with
manual launching of each investigation due to the relatively limited number of
 currently detected events (and consequently a relatively limited
number of pairs). However, even at this current level, as is noted in each
of the aforementioned references, the entire set of multiplets needed for a
complete lensing analysis would prove to be too computationally intensive to
analyse fully given the relatively high computational cost of performing them.

The need to avoid this problem in the prior investigations has motivated the
development of a number of higher speed but less complete investigations that
are able to serve as filters to limit the number of candidate multiplets that
require investigation. In general, these analyse the similarity between
different subsets of information generated under the unlensed hypotheses. These
will support lensed candidates, however, are also unable to rule out
co-incidental overlap of unlensed events maintaining the ultimate need for
\gls{pe} based investigations in these cases. We will now describe briefly each
of these analyses.

The most broad of these analyses is that proposed
by~\cite{haris_posterior_overlap} implemented in the \textsc{posterior\_overlap}
pipeline in which the posteriors obtained from standard unlensed \gls{pe}
analysis may be compared through the computation of a data-driven Bayes factor
given by
\begin{equation}
  \label{eq:posterior-overlap-bayes-factor}
  \mathcal{B}^{\textrm{L}}_{\textrm{U}} = \int
  \frac{P\left(\bm{\Theta} | d_{1}\right) P\left(\bm{\Theta} | d_{2}\right)}
       {P(\bm{\Theta})} 
  d\bm{\Theta},
\end{equation}
where $\bm{\Theta}$ is a subset of the parameters estimated in unlensed
analysis. To improve the efficiency of this
statistic,~\cite{haris_posterior_overlap} proposed to also include information
about the distributions of time delays expected under the unlensed and lensed
hypotheses i.e.
\begin{equation}
  \label{eq:rlu}
  \mathcal{R}^{\textrm{gal}} =
  \frac{P\left(\Delta{t} | \mathcal{H}_{\textrm{SL}} \right)}
       {P\left(\Delta{t} | \mathcal{H}_{\textrm{UL}} \right)}. 
\end{equation}
\cite{more_statistics} then refined this proposition further to include the
remaining effective lensing parameters---the relative magnification and Morse
phase shift---to compute the statistic
\begin{equation}
  \label{eq:mgal}
  \mathcal{M}^{\textrm{gal}} = 
  \frac{P\left(
    \Delta{t}, \mu_{r}, \Delta{\psi} | \mathcal{H}_{\textrm{SL}}
  \right)}
       {P\left(
    \Delta{t}, \mu_{r}, \Delta{\psi} | \mathcal{H}_{\textrm{UL}}
  \right)}.
\end{equation}
These may be computed from extensive catalogues of lensed and unlensed
simulations to build the distributions. Using such catalogues, the
\textsc{atlenstics} pipeline was built as a thin wrapper for the production of
these statistics from the posterior samples from unlensed \gls{pe}
investigation. The inclusion of simulations for the production of lensed
catalogues, however, does mean that these statistics are only valid under the
assumptions used to generate the populations used to create them. 

Moving from the complete set of the posteriors,~\cite{phazap}---implemented in
the \textsc{phazap} pipeline---proposed to perform an analysis of specifically
the phase posterior distributions obtained after post-processing the samples to
establish a common reference frame. Under the assumption that the posteriors on
these detected phases are sufficiently Gaussian, one may quantify the
consistency between the posteriors through a distance given by
\begin{equation}
  \label{eq:phazap-posterior-distance}
  D_{12} = \sqrt{
    \Delta{\theta}^{T} \left(C_{1} + C_{2}\right)^{-1} \Delta{\theta}
  },
\end{equation}
where $\Delta{\theta}$ is the difference between the parameters for each
individual event and the $C_{i}$ terms indicate the covariances. Computing this
quantity over the set of possible orderings and the possible phase differences
and maximising the choice from the former and minimising from the latter gives
a metric by which to assess pairs. This may be matched with a statistic to
measure how well constrained the parameters are to yield an overall ranking of
candidates to assess which ones would warrant further analysis with the
\gls{pe}-based analyses.

Each of the previous analyses has required the production of full unlensed
\gls{pe}-based analyses in order to assess whether or not a given pair would
appear to be consistent with the lensing hypothesis introducing a significant
latency between the detection of a \gls{gw} event and the ability to perform
the aforementioned analyses. To circumvent this~\cite{goyal_lensid} proposed,
and implemented in the \textsc{LensID} pipeline, to use some of the first
pieces of information produced in the \gls{gw} analysis chain---time-frequency
maps, also known as Q-transforms, which may be produced directly from incoming
data and localization skymaps which are produced rapidly using
\textsc{Bayestar}~\citep{Singer:2015ema} to identify candidates. They do this
by using both of these aforementioned pieces of information as inputs to a
machine-learning classifier which combines both intrinsic and extrinsic
parameter information encoded in these inputs to identify potential candidates.

\subsection{Automated Workflow}\label{subsec:lensingflow-workflow}

\begin{figure}
  \centering

  \caption{
    Flowchart summarising the automated workflow implemented by
    \textsc{LensingFlow} including both the multiplet and single event
    analyses. We note that the latter stages noting the output to
    \textsc{CBCFlow} happen repeatedly throughout the process of the deployment
    of the analyses such that the metadata is always reflective of the current
    state of analyses. We also note that whilst the status of the event is
    always checked after each analysis for interest in the event/pair for that
    pipeline, the status recorded in the ledger, and ultimately, the outputs,
    we show this step only for those cases where that check has knock on
    effects in the deployment of other analyses.
  }
  \label{fig:lf-flowchart}

  \begin{tikzpicture}
  \node (unlensed-analyses) [function, align=center] {
      Unlensed analyses
  };

  \node (original-metadata) [
    io, below=0.5cm of unlensed-analyses, align=center
  ] {
      \textsc{CBCFlow} metadata
  };

  \node (meets-lensing-criteria) [
    decision, below=0.5cm of original-metadata, align=center
  ] {
      Meets criteria for analysis?
  };

  \node (asimov-ledger) [
    io, below=0.5cm of meets-lensing-criteria, align=center
  ] {
    \textsc{Asimov} ledger
  };

  \node (single-image-analyses) [
    function, below left=0.5cm and 0.9cm of asimov-ledger, align=center
  ] {
    Single events
  };

  \node (multiplet-analyses) [
    function, below right=0.5cm and 0.9cm of asimov-ledger, align=center
  ] {
    Multiplets
  };

  \node (gravelamps-point-mass) [
    function, below=0.5cm of single-image-analyses, align=center
  ] {
    \textsc{Gravelamps} \\
    point mass \\
    model
  };

  \node (gravelamps-milli) [
    function, 
    below=0.5cm of single-image-analyses,
    right=0.25cm of gravelamps-point-mass,
    align=center
  ] {
    \textsc{Gravelamps} \\
    unmodelled
  };

  \node (golum-type-ii) [
    function,
    below=0.5cm of single-image-analyses,
    left=0.25cm of gravelamps-point-mass,
    align=center
  ] {
    \textsc{Golum} \\
    Type II
  };

  \node (low-latency-data) [
    function,
    below=0.5cm of multiplet-analyses,
    align=center
  ] {
    Low latency \\
    analyses \\
    \textsc{Fast Golum} \\
    \textsc{LensID} \\
    \textsc{PO} \\
    \textsc{Phazap}
  };

  \node (atlenstics) [
    function,
    below=0.5cm of multiplet-analyses,
    right=0.25cm of low-latency-data,
    align=center
  ] {
    Galaxy lens \\ 
    prior\\ compatibility \\
    \textsc{Atlenstics}
  };

  \node (ml-is-interesting) [
    decision,
    below=0.5cm of gravelamps-point-mass,
    align=center
  ] {
    Above pipeline\\
    interest threshold?
  };

  \node (sis) [
    function,
    below=0.5cm of ml-is-interesting,
    align=center
  ] {
    \textsc{Gravelamps} \\
    SIS model
  };

  \node (ll-is-interesting) [
    decision,
    below left=0.5cm and -1.3cm of low-latency-data,
    align=center
  ] {
    Above pipeline\\
    interest threshold?
  };

  \node (atlenstics-is-interesting) [
    decision,
    below=0.5cm of atlenstics,
    right=0.2cm of ll-is-interesting,
    align=center
  ] {
    Above pipeline\\
    interest threshold?
  };

  \node (priority) [
    function,
    below=0.5cm of atlenstics-is-interesting,
    align=center
  ] {
    Add priority to \\
    high latency \\
    analyses
  };
  
  \node (hl-started) [
    decision,
    below=0.5cm of ll-is-interesting,
    left=0.5cm of priority,
    align=center
  ] {
    High latency \\
    analyses started?
  };

  \node (hl-threshold) [
    decision,
    below=0.5cm of hl-started,
    align=center
  ] {
    Above threshold \\ for two analyses?
  };

  \node (hl-analyses) [
    function,
    below=0.5cm of hl-threshold,
    align=center
  ] {
    Joint \gls{pe} \\
    analyses \\
    \textsc{Golum} \\
    \textsc{Hanabi}
  };

  \node (final-ledger) [
    io,
    below=10.5cm of asimov-ledger,
    align=center
  ] {
    \textsc{Asimov} ledger
  };

  \node (final-metadata) [
    io,
    below left=0.5cm and 0.75cm of final-ledger,
    align=center
  ] {
    \textsc{CBCFlow} \\ metadata
  };

  \node (html) [
    io,
    below right=0.47cm and 0.75cm of final-ledger,
    align=center
  ] {
    HTML \\ Summary Page
  };

  \draw [arrow] (unlensed-analyses) -- (original-metadata);
  \draw [arrow] (original-metadata) -- (meets-lensing-criteria);
  \draw [arrow] 
    (meets-lensing-criteria) -- (asimov-ledger)
    node[anchor=west, pos=0.5]{Y};
  \draw [arrow] (asimov-ledger) -| (single-image-analyses);
  \draw [arrow] (asimov-ledger) -| (multiplet-analyses);
  \draw [arrow] (single-image-analyses) -- (gravelamps-point-mass);
  \draw [arrow] (single-image-analyses) -| (gravelamps-milli);
  \draw [arrow] (single-image-analyses) -| (golum-type-ii);
  \draw [arrow] (multiplet-analyses) -- (low-latency-data);
  \draw [arrow] (multiplet-analyses) -| (atlenstics);
  \draw [arrow] (gravelamps-point-mass) -- (ml-is-interesting);
  \draw [arrow] 
    (ml-is-interesting) -- (sis)
    node[anchor=east, pos=0.5]{Y};
  \draw [arrow] 
    (low-latency-data.south -| ll-is-interesting) -- (ll-is-interesting);
  \draw [arrow] 
    (atlenstics.south -| atlenstics-is-interesting) --
    (atlenstics-is-interesting);
  \draw [arrow]
    (ll-is-interesting.south -| hl-started) -- (hl-started)
    node[anchor=east, pos=0.7]{Y};
  \draw [arrow]
    (atlenstics-is-interesting.south -| priority) -- (priority)
    node[anchor=west, pos=0.7]{Y};
  \draw [arrow]
    (hl-started) -- (priority)
    node[anchor=south, pos=0.7]{Y};
  \draw [arrow]
    (hl-started.south -| hl-threshold) -- (hl-threshold)
    node[anchor=east, pos=0.7]{N};
  \draw [arrow]
    (hl-threshold) -- (hl-analyses)
    node[anchor=east, pos=0.7]{Y};
  \draw [arrow] (priority) |- (hl-analyses);
  \draw [arrow] (golum-type-ii) |- (final-ledger);
  \draw [arrow] (sis) |- (final-ledger);
  \draw [arrow] (gravelamps-milli) -- (final-ledger);
  \draw [arrow] (hl-analyses) -| (final-ledger);
  \draw [arrow] (final-ledger) |- (final-metadata);
  \draw [arrow] (final-ledger) |- (html);

\end{tikzpicture}
\end{figure}
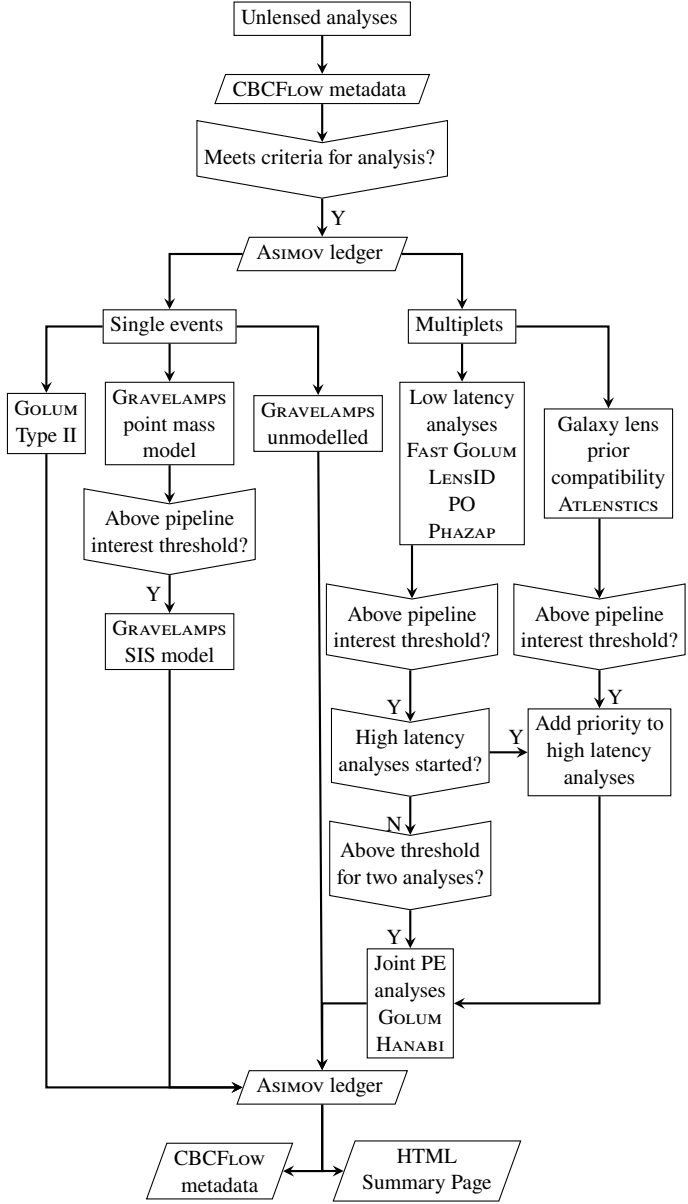

We now describe in detail the construction of the automated workflow and the
necessary implementations for deploying it in the manner described. A summary
of the workflow is provided in Figure~\ref{fig:lf-flowchart}. We note that to
prepare for deployment in the automated workflow, each pipeline was updated to
include the necessary implementation of the \textsc{Pipeline}
class~\citep{asimov_docs} that makes it usable by the \textsc{Asimov}
framework~\citep{asimov}. Each pipeline was then able to produce a standard
template of their respective configuration file for which \textsc{Asimov} is
then able to fill with run-specific information. 

The initial stage of work for the production of lensing analyses under the
automated workflow is gathering the requisite \textsc{CBCFlow} metadata files
into a library of events to be analysed~\citep{cbcflow}. Under general
deployment conditions lensing analyses may be considered as part of a larger
chain of analyses from initial detection through to unlensed \gls{pe}, each of
which may contribute to a large \textsc{CBCFlow} library containing both
\gls{gw} events as well as transient noise features, so-called ``glitches''.
The first step implemented is a means to filter events through a selection
criteria for lensing analyses. For this work, the selection
criteria are that the event is determined to most likely be a binary black hole
merger and that the event has been detected by search pipelines, such as
\textsc{gstLAL}~\citep{Messick:2016aqy, Sachdev:2019vvd, CANNON2021100680,
Tsukada:2023edh} or \textsc{PyCBC}~\citep{Biwer:2018osg, Allen:2005fk,
Allen:2004gu, Nitz:2017svb, DalCanton:2014hxh}, that the \gls{far} is less than
$1$ per year. Events passing these criteria are transferred to an individual
sub-library for the lensing analyses.

Once the library of metadata is constructed, the next prepatory step is using
the information contained within to build an \textsc{Asimov} ledger which
serves as the central hub for that framework to automate the deployment and
monitoring of the analyses of the events and consequent pairs. For pipelines,
such as \textsc{CBCFlow}, that are compatible with \textsc{Asimov}, this is
achieved through a standard \textsc{Applicator} class implemented within the
\textsc{Asimov} codebase~\citep{asimov_docs}. Whilst this existed already for
\textsc{CBCFlow} and would successfully gather the information needed to produce
unlensed analyses, modifications were needed to gather additional information
from those unlensed analyses for the purposes of lensing analyses. These
modifications are made available in \textsc{Asimov} directly in releases
beginning with the $0.6$ series. 

With the events applied to the ledger, the analysis configurations must be set
up for both those analyses that use only that single event as well as those
analyses that cover multiplets of events. These are applied using a
\textsc{YAML} configuration file that describes essential properties of each
analysis. The \textsc{LensingFlow} package produces this file automatically
when applying new events and it will generate all necessary individual
analyses, as well as both the pairs formed by the newly applied events and
those from the new events with those events that are already in the ledger
whilst avoiding duplication through the redundant other ordering of each
pair---specifically by ensuring that all multiplet-based analyses are done in a
time-of-detection ordering and that relevant information about the unlensed
analysis configurations and results are retrieved from relevant files linked in
the \textsc{CBCFlow} metadata.

The \textsc{Asimov} monitor will then use the information given from both the
event application as well as the analyses' configurations to build and submit
the requested jobs for the \textsc{HTCondor} scheduler~\citep{condor-hunter}
widely used on the cluster computing facilities necessary to facilitate high
volume analysis of \gls{gw} signal data. It will begin with the low latency
analyses for multiplets---in this work this consists of \textsc{LensID},
\textsc{Posterior\_Overlap}, \textsc{Phazap}, and \textsc{Golum} operating a
conditional pair-wise \gls{pe} approach---and the check for compatibility with
expectations for galaxy lensing using \textsc{atlenstics}, as well as the
individual event analysis consisting of the \textsc{Golum} Type II \gls{pe}
analysis, and the \textsc{Gravelamps} modelled and unmodelled \gls{pe}. Upon
the completion of each pipeline, the results are automatically analysed to
determine if the support for lensing from each pipeline is above a user-defined
threshold for that pipeline which warrants additional investigations. 

Should a pipeline from the low latency multiplet analyses yield an
above-threshold candidate, \textsc{Asimov} will wait for additional information
before proceeding with any follow-up steps. If a second pipeline concurs,
\textsc{Asimov} will then start the joint \gls{pe} analysis pipelines
\textsc{Golum} and \textsc{Hanabi}. Additional low latency pipelines that
concur with the two sufficient for starting the high latency analyses will
provide additional relative priority in the \textsc{HTCondor} scheduler for
analyses of this multiplet. Conversely, should enough low latency pipelines
indicate a lack of support for lensing such that the starting of the higher
latency analyses is no longer possible based on the remaining pipelines,
\textsc{Asimov} will automatically discard the \gls{pe} analyses whilst
retaining the remaining low latency analyses for the record. We note here that
the \textsc{atlenstics} consistency evaluation is not used for the purposes of
determining whether the \gls{pe}-based analyses are started and is instead only
used for the purposes of prioritisation. This is to prevent the possibility of
prior belief overriding support from the data in regions that are unexpected
whilst still rewarding those multiplets that are within expectations.

For the other branches of analyses, only the isolated point mass model analyses
causes another analysis to be directly started by \textsc{Asimov} in the case
when it reports that the analysis has found an above-threshold candidate. In
this case, \textsc{Asimov} will deploy an additional lens model---the singular
isothermal sphere~\citep{Binney:1989je}---for investigation. This model, whilst
not necessarily the most reflective of objects on the physical mass scale that
results in single event behaviour, is useful in the context of looking at an
extended object vs a more point-like object. For the remaining branches, as
well as for any triggered analyses, when they conclude, the same exercise is
undertaken with the results analysed to compare to a threshold. In this case,
this is simply recorded in the ledger with no direct effects but does indicate
that the event or multiplet should undergo further manual investigations such
as those in e.g.~\cite{Janquart:2023mvf}.

During the entire process, as modifications are made to the \textsc{Asimov}
ledger, these are also fed back to the \textsc{CBCFlow} metadata through a
standard \textsc{Collector} class implemented within
\textsc{Asimov}~\citep{asimov_docs}. This ensures that these files are kept
up-to-date with a current state of analyses. Again, modifications were made
here to properly transfer back information from the lensing analyses discussed
in this work and are made available in \textsc{Asimov} beginning with the $0.6$
series. Additionally, the \textsc{LensingFlow} package provides automated
production of an HTML page to monitor the current state of each analysis in the
ledger and provides lists of those jobs that have passed the
thresholds for interest and those jobs that have encountered technical
difficulties causing them to get stuck, allowing an at-a-glance status
update for lensing analyses.

\section{Example Deployment}\label{sec:deployment}
To provide an initial proof-of-concept deployment of the \textsc{LensingFlow},
it was applied to a \gls{mdc} comprising $16$ \gls{gw} signals that were
injected into simulated Gaussian noise drawn from the expected noise \gls{psd}
for the currently ongoing fourth observing run~\citep{o4_psd}. The parameters
from these signals were drawn from a \textsc{PowerLaw+Peak} merger population
model---the parameters of which were themselves drawn from the posterior
samples fitting this model to the population of observed binaries from GWTC-3
in~\cite{o3_randp}. Data was generated for the span of $\sim$1 month and
assuming no detector down-time in a network consisting of the two
\glsfmtshort{ligo} detectors and the Virgo detector. Generated systems were
then also randomly assigned lensing effects. The signals were produced using
the \textsc{IMRPhenomXPHM} waveform approximant~\citep{Pratten:2020ceb}. 

Analysis was performed to mirror the detection and catalogue unlensed \gls{pe}
stages for these signals. Of the total number of injected signals, $10$ were
found by the \textsc{gstLAL} detection pipeline and were thus subjected to
\gls{pe} analysis and made available to the \textsc{LensingFlow} for analysis
using \textsc{CBCFlow} metadata files generated from the unlensed analysis
results. A summary of the events is provided in Table~\ref{tab:mdc-events}.

\begin{table}
  \centering

  \caption{Summary of the detected events used in the \gls{mdc} to which the
    \textsc{LensingFlow} was applied as a proof-of-concept demonstration. Shown
    are the name of the event, the individual source frame masses of the
    primary and secondary, apparent luminosity distance to the system, and the
    kind(s) of lensing, if any, applied to the event. We note that for Multiplet
    images (M), the image type is noted. PMS refers to point mass singlet and
    US to unmodelled singlet respectively. Superscripts identify events
    belonging to the same lensed system.}
  \label{tab:mdc-events}

  \begin{tabularx}{\columnwidth}{cXXXX}

    \hline \hline Event & Primary Mass $(M_{\odot})$ & Secondary Mass
    $(M_{\odot})$ & Apparent Luminosity Distance $(\textrm{Mpc})$ & Lensing
    Applied \\

    \hline MS220425h & $35.5$ & $33.9$ & $5435.0$ & None \\
    MS220505y & $21.2$ & $19.2$ & $1581.8$ & None \\
    MS220508a$^{1}$ & $70.1$ & $38.8$ & $2858.2$ & Type I M + PMS \\
    MS220508b$^{1}$ & $70.1$ & $38.8$ & $3366.5$ & Type II M \\
    MS220509x$^{1}$ & $70.1$ & $38.8$ & $5960.8$ & Type II M \\
    MS220510af$^{2}$ & $166.9$ & $117.0$ & $9250.3$ & Type I M \\
    MS220510ae$^{2}$ & $166.9$ & $117.0$ & $10066.8$ & Type II M \\
    MS220514y & $77.8$ & $57.1$ & $7324.3$ & None \\
    MS220523d & $10.2$ & $9.3$ & $1935.2$ & None \\
    MS220526y & $56.1$ & $53.9$ & $3579.4$ & US \\
    \hline

  \end{tabularx}

\end{table}

We will now briefly summarise the results from those analyses that are able to
start downstream analyses based on their results. Beginning with the pair-wise
analyses the \textsc{Phazap} pipeline identified $6$ pairs to be passed to high
latency, the \textsc{Posterior\_Overlap} pipeline identified $3$, the
\textsc{LensID} passed $4$, and the \textsc{golum} pipeline operating in a
conditional approach also identified $6$ pairs. A summary of these pairs is
provided in Table~\ref{tab:low-latency-results}. From the $7$ unique pairs
identified from these analyses, $6$ are above threshold in two or more
pipelines and would be passed to the joint \gls{pe} analyses---which include
those that result from the three combinations from the triplet
$08$a-$08$b-$09$x and the pair $10$ae-$10$af which were the injected lensed
cases noted in Table~\ref{tab:mdc-events}. All other pairs not listed were
discarded. This phase would therefore narrow the total number of pairs that
would require analysis from $45$ to $6$, a $\sim85$\% decline in computational
burden. 

Turning to the isolated point mass analysis, \textsc{Gravelamps} identified
both the events MS$220508$a and MS$220526$y as worthy of additional
investigation---corresponding to the injections with single image lensing
applied. That the isolated point mass investigation was able to identify both
the modelled and unmodelled events may indicate that a further optimisation
that could be made to decrease computational burden would be to analyse events
with only this model as a first step. However, further investigations would
need to be carried out for comparative analysis between signals generated from
the modelled and unmodelled approaches.

\begin{table}
  \centering

  \caption{Summary of the pairs identified by the low latency pair-wise
  analyses and which analyses found them.}
  \label{tab:low-latency-results}

  \begin{tabularx}{\columnwidth}{cXXXXX}

    \hline \hline Pair & \textsc{Fast Golum} & \textsc{LensID} &
    \textsc{Phazap} & \textsc{Posterior Overlap} \\
    
    \hline MS220508a \& MS220508b & Y & Y & Y & Y \\
    MS220508a \& MS220509x & Y & Y & Y & Y \\
    MS220508b \& MS220509x & Y & N & Y & N \\
    MS220510ae \& MS220510af & Y & Y & Y & Y \\
    MS220425h \& MS220510af & N & N & Y & N \\
    MS220425h \& MS220510ae & Y & N & Y & N \\
    MS220510ae \& MS220514y & Y & Y & N & N \\ 
    \hline

  \end{tabularx}

\end{table}

Whilst the remainder of the analyses do not directly contribute to any of the
\textsc{LensingFlow} functionality directly, we note for completeness that in
terms of the automated thresholds which causes \textsc{LensingFlow} to note in
the metadata that an event/pair requires further investigation---the Type II
strong lensing analysis flagged the events MS$220508$a and MS$220425$h, and the
joint \gls{pe} analyses flagged all $4$ pairs from the lensed events as worthy
of additional scrutiny.

From this we may see that the automated operation of the pipelines is able to
successfully filter to those candidates that have lensing signatures and start
\gls{pe} based analyses, allowing the continued computational burden reduction
that each of these pipelines offer. Aside from the initial input of the
\textsc{CBCFlow} metadata files, the operation of these investigations required
no manual intervention. In a scaled deployment this would significantly free-up
analyst time to devote to specific follow-ups or other investigations allowing
more detailed lensing investigations over the course of e.g. an observing run.

\section{Conclusion}\label{sec:conclusion}
In the coming years, as additional \gls{gw} detectors come online and the
sensitivities of the detectors continues to improve, both the number and rate
of \gls{gw} detections will continue to rise. It is, therefore, imperative that
robust infrastructure is in place to automate currently manually performed
analyses. Otherwise, the workload will eventually exceed what can be manually
handled in a reasonable amount of time. The need for such automation is
particularly strongly felt in analyses for \gls{gw} lensing in which multiplets
of events must be analysed, significantly increasing the workload as compared
to non-lensed single event analyses. Additionally, lensing pipelines represent
a relatively intricate workflow compared to standard chains of analyses, or
sets of non-interdependent analyses which are common in \gls{gw} analyses,
marking a more unique use case for the steps towards automated analysis taken
in the community with the development of tools such as
\textsc{CBCFlow}~\citep{cbcflow} and \textsc{Asimov}~\citep{asimov}.

To address these needs of the \gls{gw} lensing community, in this work we have
presented \textsc{LensingFlow}---an automated workflow for lensing analyses
using at its base a series of customised modifications to the two
aforementioned frameworks as well as custom functionality encoded in a
\textsc{python} package, that we make available freely alongside this
work~\citep{lensingflow}. This framework provides the means to perform analyses
both of individual \gls{gw} events as well as the pairwise combinations
necessary for a full examination of the lensing hypothesis across the different
regimes. Currently the framework includes a number of flagship lensing analysis
pipelines, but the framework is flexible and able to accommodate additional
pipelines to ensure that the needs of the community may continue to be met by
it in the future. For example, work has already started for
\textsc{LensingFlow} to handle inclusions in \textsc{CBCFlow} metadata files
from candidates for searches from sub-threshold searcehs for faint additional
lensed images not included in the original GWTC selections, though those
pipelines~\citep{McIsaac:2019use, Li:2019osa, Li:2023zdl} have not been fully
integrated into the workflow yet.

We demonstrate an initial proof-of-concept deployment on an \gls{mdc}
comprising a set of 10 events including examples of each of the regimes of
lensing that analyses consider, under the automated passing criteria, the
workflow correctly both generated the more in-depth analyses for the lensed
events as well as identifying them for further scrutiny as well as reducing the
workload of the most computationally intensive investigations of multiple
events by $\sim85$\%. The automated in-fill of information also prevents the
development of inconsistencies between the multiple analyses of the same
event/multiplet, ensuring that comparisons between the analyses are
always apples-to-apples whilst minimising the human effort required to achieve
these needs, which in a realistic deployment would free analyst time for
further investigations of any high priority candidates, and other scientific
work.

\section*{Acknowledgements}\label{sec:acknowledgements}
The authors are grateful to Tomasz Baka for careful review of the paper. The
authors are grateful for computational resources provided by the LIGO
Laboratory and supported by National Science Foundation Grants PHY-0757058 and
PHY-0823459. MW is supported by the research programme of the Netherlands
Organisation for Scientific Research (NWO). PC and DK were supported by the
Universitat de les Illes Balears (UIB); the Spanish Agencia Estatal de
Investigaci{\'o}n grants CNS2022-135440, PID2022-138626NB-I00,
RED2024-153978-E, RED2024-153735-E, funded by MICIU/AEI/10.13039/501100011033,
the European Union NextGenerationEU/PRTR, and the ERDF/EU; and the Comunitat
Aut{\'o}noma de les Illes Balears through the Conselleria d'Educaci{\'o} i
Universitats with funds from the European Union--NextGenerationEU/PRTR-C17.I1
(SINCO2022/6719) and from the European Union--European Regional Development
Fund (ERDF) (SINCO2022/18146). ES is supported by grants from the College of
Science and Engineering at the University of Glasgow. LEU is supported by the
Hong Kong PhD Fellowship Scheme (HKPFS) from the Hong Kong Research Grants
Council (RGC). OAH and LEU acknowledge suport by grants from the Research
Grants Council of Hong Kong (Project No. CUHK 14304622, 14307923, and
14307724), the start-up grant from the Chinese University of Hong Kong, and the
Direct Grant for Research from the Research Committee of The Chinese University
of Hong Kong. JCLC, JME, and RKLL are supported by VILLUM FONDEN (grant no.
53101 and 37766). The Center of Gravity is a Center of Excellence funded by the
Danish National Research Foundation under grant No. 184. The Tycho
supercomputer hosted at the SCIENCE HPC center at the Univesrity of Copenhagen
was used for supporting this work. JME is also supported by the European
Union's Horizon 2020 research and innovation program under the Marie
Sk\l{}odowska-Curie grant agreement No. 847523 INTERACTIONS. PP would like to
thank Prof. Sukanta Bose for his support and IUCCA, Pune for providing
computational facilities.

\bibliographystyle{aasjournal}
\bibliography{lensingflow.bib}

\begin{thebibliography}{}
\expandafter\ifx\csname natexlab\endcsname\relax\def\natexlab#1{#1}\fi
\providecommand{\url}[1]{\href{#1}{#1}}
\providecommand{\dodoi}[1]{doi:~\href{http://doi.org/#1}{\nolinkurl{#1}}}
\providecommand{\doeprint}[1]{\href{http://ascl.net/#1}{\nolinkurl{http://ascl.net/#1}}}
\providecommand{\doarXiv}[1]{\href{https://arxiv.org/abs/#1}{\nolinkurl{https://arxiv.org/abs/#1}}}

\bibitem[{Aasi {et~al.}(2015)}]{ligo_detector}
Aasi, J., {et~al.} 2015, Class. Quant. Grav., 32, 074001,
  \dodoi{10.1088/0264-9381/32/7/074001}

\bibitem[{Abbott {et~al.}(2016)Abbott, Abbott, Abbott, Abernathy, Acernese,
  Ackley, Adams, Adams, Addesso, Adhikari, Adya, Affeldt, Agathos, Agatsuma,
  Aggarwal, Aguiar, Aiello, Ain, Ajith, Allen, Allocca, Altin, Anderson,
  Anderson, Arai, Arain, Araya, Arceneaux, Areeda, Arnaud, Arun, Ascenzi,
  Ashton, Ast, Aston, Astone, Aufmuth, Aulbert, Babak, Bacon, Bader, Baker,
  Baldaccini, Ballardin, Ballmer, Barayoga, Barclay, Barish, Barker, Barone,
  Barr, Barsotti, Barsuglia, Barta, Bartlett, Barton, Bartos, Bassiri, Basti,
  Batch, Baune, Bavigadda, Bazzan, Behnke, Bejger, Belczynski, Bell, Bell,
  Berger, Bergman, Bergmann, Berry, Bersanetti, Bertolini, Betzwieser, Bhagwat,
  Bhandare, Bilenko, Billingsley, Birch, Birney, Birnholtz, Biscans, Bisht,
  Bitossi, Biwer, Bizouard, Blackburn, Blair, Blair, Blair, Bloemen, Bock,
  Bodiya, Boer, Bogaert, Bogan, Bohe, Bojtos, Bond, Bondu, Bonnand, Boom, Bork,
  Boschi, Bose, Bouffanais, Bozzi, Bradaschia, Brady, Braginsky, Branchesi,
  Brau, Briant, Brillet, Brinkmann, Brisson, Brockill, Brooks, Brown, Brown,
  Brown, Buchanan, Buikema, Bulik, Bulten, Buonanno, Buskulic, Buy, Byer,
  Cabero, Cadonati, Cagnoli, Cahillane, Bustillo, Callister, Calloni, Camp,
  Cannon, Cao, Capano, Capocasa, Carbognani, Caride, Diaz, Casentini, Caudill,
  Cavagli\`a, Cavalier, Cavalieri, Cella, Cepeda, Baiardi, Cerretani, Cesarini,
  Chakraborty, Chalermsongsak, Chamberlin, Chan, Chao, Charlton,
  Chassande-Mottin, Chen, Chen, Cheng, Chincarini, Chiummo, Cho, Cho, Chow,
  Christensen, Chu, Chua, Chung, Ciani, Clara, Clark, Cleva, Coccia, Cohadon,
  Colla, Collette, Cominsky, Constancio, Conte, Conti, Cook, Corbitt, Cornish,
  Corsi, Cortese, Costa, Coughlin, Coughlin, Coulon, Countryman, Couvares,
  Cowan, Coward, Cowart, Coyne, Coyne, Craig, Creighton, Creighton, Cripe,
  Crowder, Cruise, Cumming, Cunningham, Cuoco, Canton, Danilishin, D'Antonio,
  Danzmann, Darman, Da~Silva~Costa, Dattilo, Dave, Daveloza, Davier, Davies,
  Daw, Day, De, DeBra, Debreczeni, Degallaix, De~Laurentis, Del\'eglise,
  Del~Pozzo, Denker, Dent, Dereli, Dergachev, DeRosa, De~Rosa, DeSalvo,
  Dhurandhar, D\'{\i}az, Di~Fiore, Di~Giovanni, Di~Lieto, Di~Pace, Di~Palma,
  Di~Virgilio, Dojcinoski, Dolique, Donovan, Dooley, Doravari, Douglas, Downes,
  Drago, Drever, Driggers, Du, Ducrot, Dwyer, Edo, Edwards, Effler, Eggenstein,
  Ehrens, Eichholz, Eikenberry, Engels, Essick, Etzel, Evans, Evans, Everett,
  Factourovich, Fafone, Fair, Fairhurst, Fan, Fang, Farinon, Farr, Farr,
  Favata, Fays, Fehrmann, Fejer, Feldbaum, Ferrante, Ferreira, Ferrini,
  Fidecaro, Finn, Fiori, Fiorucci, Fisher, Flaminio, Fletcher, Fong, Fournier,
  Franco, Frasca, Frasconi, Frede, Frei, Freise, Frey, Frey, Fricke, Fritschel,
  Frolov, Fulda, Fyffe, Gabbard, Gair, Gammaitoni, Gaonkar, Garufi, Gatto,
  Gaur, Gehrels, Gemme, Gendre, Genin, Gennai, George, Gergely, Germain, Ghosh,
  Ghosh, Ghosh, Giaime, Giardina, Giazotto, Gill, Glaefke, Gleason, Goetz,
  Goetz, Gondan, Gonz\'alez, Castro, Gopakumar, Gordon, Gorodetsky, Gossan,
  Gosselin, Gouaty, Graef, Graff, Granata, Grant, Gras, Gray, Greco, Green,
  Greenhalgh, Groot, Grote, Grunewald, Guidi, Guo, Gupta, Gupta, Gushwa,
  Gustafson, Gustafson, Hacker, Hall, Hall, Hammond, Haney, Hanke, Hanks,
  Hanna, Hannam, Hanson, Hardwick, Harms, Harry, Harry, Hart, Hartman, Haster,
  Haughian, Healy, Heefner, Heidmann, Heintze, Heinzel, Heitmann, Hello,
  Hemming, Hendry, Heng, Hennig, Heptonstall, Heurs, Hild, Hoak, Hodge, Hofman,
  Hollitt, Holt, Holz, Hopkins, Hosken, Hough, Houston, Howell, Hu, Huang,
  Huerta, Huet, Hughey, Husa, Huttner, Huynh-Dinh, Idrisy, Indik, Ingram, Inta,
  Isa, Isac, Isi, Islas, Isogai, Iyer, Izumi, Jacobson, Jacqmin, Jang, Jani,
  Jaranowski, Jawahar, Jim\'enez-Forteza, Johnson, Johnson-McDaniel, Jones,
  Jones, Jonker, Ju, Haris, Kalaghatgi, Kalogera, Kandhasamy, Kang, Kanner,
  Karki, Kasprzack, Katsavounidis, Katzman, Kaufer, Kaur, Kawabe, Kawazoe,
  K\'ef\'elian, Kehl, Keitel, Kelley, Kells, Kennedy, Keppel, Key,
  Khalaidovski, Khalili, Khan, Khan, Khan, Khazanov, Kijbunchoo, Kim, Kim, Kim,
  Kim, Kim, Kim, King, King, Kinzel, Kissel, Kleybolte, Klimenko, Koehlenbeck,
  Kokeyama, Koley, Kondrashov, Kontos, Koranda, Korobko, Korth, Kowalska,
  Kozak, Kringel, Krishnan, Kr\'olak, Krueger, Kuehn, Kumar, Kumar, Kuo,
  Kutynia, Kwee, Lackey, Landry, Lange, Lantz, Lasky, Lazzarini, Lazzaro,
  Leaci, Leavey, Lebigot, Lee, Lee, Lee, Lee, Lenon, Leonardi, Leong, Leroy,
  Letendre, Levin, Levine, Li, Libson, Littenberg, Lockerbie, Logue, Lombardi,
  London, Lord, Lorenzini, Loriette, Lormand, Losurdo, Lough, Lousto, Lovelace,
  L\"uck, Lundgren, Luo, Lynch, Ma, MacDonald, Machenschalk, MacInnis, Macleod,
  Maga\~na Sandoval, Magee, Mageswaran, Majorana, Maksimovic, Malvezzi, Man,
  Mandel, Mandic, Mangano, Mansell, Manske, Mantovani, Marchesoni, Marion,
  M\'arka, M\'arka, Markosyan, Maros, Martelli, Martellini, Martin, Martin,
  Martynov, Marx, Mason, Masserot, Massinger, Masso-Reid, Matichard, Matone,
  Mavalvala, Mazumder, Mazzolo, McCarthy, McClelland, McCormick, McGuire,
  McIntyre, McIver, McManus, McWilliams, Meacher, Meadors, Meidam, Melatos,
  Mendell, Mendoza-Gandara, Mercer, Merilh, Merzougui, Meshkov, Messenger,
  Messick, Meyers, Mezzani, Miao, Michel, Middleton, Mikhailov, Milano, Miller,
  Millhouse, Minenkov, Ming, Mirshekari, Mishra, Mitra, Mitrofanov,
  Mitselmakher, Mittleman, Moggi, Mohan, Mohapatra, Montani, Moore, Moore,
  Moraru, Moreno, Morriss, Mossavi, Mours, Mow-Lowry, Mueller, Mueller, Muir,
  Mukherjee, Mukherjee, Mukherjee, Mukund, Mullavey, Munch, Murphy, Murray,
  Mytidis, Nardecchia, Naticchioni, Nayak, Necula, Nedkova, Nelemans, Neri,
  Neunzert, Newton, Nguyen, Nielsen, Nissanke, Nitz, Nocera, Nolting,
  Normandin, Nuttall, Oberling, Ochsner, O'Dell, Oelker, Ogin, Oh, Oh, Ohme,
  Oliver, Oppermann, Oram, O'Reilly, O'Shaughnessy, Ott, Ottaway, Ottens,
  Overmier, Owen, Pai, Pai, Palamos, Palashov, Palomba, Pal-Singh, Pan, Pan,
  Pankow, Pannarale, Pant, Paoletti, Paoli, Papa, Paris, Parker, Pascucci,
  Pasqualetti, Passaquieti, Passuello, Patricelli, Patrick, Pearlstone,
  Pedraza, Pedurand, Pekowsky, Pele, Penn, Perreca, Pfeiffer, Phelps, Piccinni,
  Pichot, Pickenpack, Piergiovanni, Pierro, Pillant, Pinard, Pinto, Pitkin,
  Poeld, Poggiani, Popolizio, Post, Powell, Prasad, Predoi, Premachandra,
  Prestegard, Price, Prijatelj, Principe, Privitera, Prix, Prodi, Prokhorov,
  Puncken, Punturo, Puppo, P\"urrer, Qi, Qin, Quetschke, Quintero,
  Quitzow-James, Raab, Rabeling, Radkins, Raffai, Raja, Rakhmanov, Ramet,
  Rapagnani, Raymond, Razzano, Re, Read, Reed, Regimbau, Rei, Reid, Reitze,
  Rew, Reyes, Ricci, Riles, Robertson, Robie, Robinet, Rocchi, Rolland,
  Rollins, Roma, Romano, Romano, Romanov, Romie, Rosi\ifmmode~\acute{n}\else
  \'{n}\fi{}ska, Rowan, R\"udiger, Ruggi, Ryan, Sachdev, Sadecki, Sadeghian,
  Salconi, Saleem, Salemi, Samajdar, Sammut, Sampson, Sanchez, Sandberg,
  Sandeen, Sanders, Sanders, Sassolas, Sathyaprakash, Saulson, Sauter, Savage,
  Sawadsky, Schale, Schilling, Schmidt, Schmidt, Schnabel, Schofield,
  Sch\"onbeck, Schreiber, Schuette, Schutz, Scott, Scott, Sellers, Sengupta,
  Sentenac, Sequino, Sergeev, Serna, Setyawati, Sevigny, Shaddock, Shaffer,
  Shah, Shahriar, Shaltev, Shao, Shapiro, Shawhan, Sheperd, Shoemaker,
  Shoemaker, Siellez, Siemens, Sigg, Silva, Simakov, Singer, Singer, Singh,
  Singh, Singhal, Sintes, Slagmolen, Smith, Smith, Smith, Smith, Son, Sorazu,
  Sorrentino, Souradeep, Srivastava, Staley, Steinke, Steinlechner,
  Steinlechner, Steinmeyer, Stephens, Stevenson, Stone, Strain, Straniero,
  Stratta, Strauss, Strigin, Sturani, Stuver, Summerscales, Sun, Sutton,
  Swinkels, Szczepa\ifmmode~\acute{n}\else \'{n}\fi{}czyk, Tacca, Talukder,
  Tanner, T\'apai, Tarabrin, Taracchini, Taylor, Theeg, Thirugnanasambandam,
  Thomas, Thomas, Thomas, Thorne, Thorne, Thrane, Tiwari, Tiwari, Tokmakov,
  Tomlinson, Tonelli, Torres, Torrie, T\"oyr\"a, Travasso, Traylor, Trifir\`o,
  Tringali, Trozzo, Tse, Turconi, Tuyenbayev, Ugolini, Unnikrishnan, Urban,
  Usman, Vahlbruch, Vajente, Valdes, Vallisneri, van Bakel, van Beuzekom,
  van~den Brand, Van Den~Broeck, Vander-Hyde, van~der Schaaf, van Heijningen,
  van Veggel, Vardaro, Vass, Vas\'uth, Vaulin, Vecchio, Vedovato, Veitch,
  Veitch, Venkateswara, Verkindt, Vetrano, Vicer\'e, Vinciguerra, Vine, Vinet,
  Vitale, Vo, Vocca, Vorvick, Voss, Vousden, Vyatchanin, Wade, Wade, Wade,
  Waldman, Walker, Wallace, Walsh, Wang, Wang, Wang, Wang, Wang, Ward, Ward,
  Warner, Was, Weaver, Wei, Weinert, Weinstein, Weiss, Welborn, Wen,
  We\ss{}els, Westphal, Wette, Whelan, Whitcomb, White, Whiting, Wiesner,
  Wilkinson, Willems, Williams, Williams, Williamson, Willis, Willke, Wimmer,
  Winkelmann, Winkler, Wipf, Wiseman, Wittel, Woan, Worden, Wright, Wu, Yablon,
  Yakushin, Yam, Yamamoto, Yancey, Yap, Yu, Yvert, Zadro\ifmmode~\dot{z}\else
  \.{z}\fi{}ny, Zangrando, Zanolin, Zendri, Zevin, Zhang, Zhang, Zhang, Zhang,
  Zhao, Zhou, Zhou, Zhu, Zucker, Zuraw, \& Zweizig}]{gw150914_discovery}
Abbott, B.~P., Abbott, R., Abbott, T.~D., {et~al.} 2016, Phys. Rev. Lett., 116,
  061102, \dodoi{10.1103/PhysRevLett.116.061102}

\bibitem[{Abbott {et~al.}(2020)Abbott, Abbott, Abbott, Abraham, Acernese,
  Ackley, Adams, Adya, Affeldt, Agathos, Agatsuma, Aggarwal, Aguiar, Aiello,
  Ain, Ajith, Akutsu, Allen, Allocca, Aloy, Altin, Amato, Ananyeva, Anderson,
  Anderson, Ando, Angelova, Antier, Appert, Arai, Arai, Arai, Araki, Araya,
  Araya, Areeda, Arène, Aritomi, Arnaud, Arun, Ascenzi, Ashton, Aso, Aston,
  Astone, Aubin, Aufmuth, AultONeal, Austin, Avendano, Avila-Alvarez, Babak,
  Bacon, Badaracco, Bader, Bae, Bae, Baiotti, Bajpai, Baker, Baldaccini,
  Ballardin, Ballmer, Banagiri, Barayoga, Barclay, Barish, Barker, Barkett,
  Barnum, Barone, Barr, Barsotti, Barsuglia, Barta, Bartlett, Barton, Bartos,
  Bassiri, Basti, Bawaj, Bayley, Bazzan, Bécsy, Bejger, Belahcene, Bell,
  Beniwal, Berger, Bergmann, Bernuzzi, Bero, Berry, Bersanetti, Bertolini,
  Betzwieser, Bhandare, Bidler, Bilenko, Bilgili, Billingsley, Birch, Birney,
  Birnholtz, Biscans, Biscoveanu, Bisht, Bitossi, Bizouard, Blackburn, Blair,
  Blair, Blair, Bloemen, Bode, Boer, Boetzel, Bogaert, Bondu, Bonilla, Bonnand,
  Booker, Boom, Booth, Bork, Boschi, Bose, Bossie, Bossilkov, Bosveld,
  Bouffanais, Bozzi, Bradaschia, Brady, Bramley, Branchesi, Brau, Briant,
  Briggs, Brighenti, Brillet, Brinkmann, Brisson, Brockill, Brooks, Brown,
  Brown, Brunett, Buikema, Bulik, Bulten, Buonanno, Buskulic, Buy, Byer,
  Cabero, Cadonati, Cagnoli, Cahillane, Bustillo, Callister, Calloni, Camp,
  Campbell, Canepa, Cannon, Cannon, Cao, Cao, Capocasa, Carbognani, Caride,
  Carney, Carullo, Diaz, Casentini, Caudill, Cavaglià, Cavalier, Cavalieri,
  Cella, Cerdá-Durán, Cerretani, Cesarini, Chaibi, Chakravarti, Chamberlin,
  Chan, Chan, Chao, Charlton, Chase, Chassande-Mottin, Chatterjee, Chaturvedi,
  Chatziioannou, Cheeseboro, Chen, Chen, Chen, Chen, Chen, Chen, Cheng, Cheong,
  Chia, Chincarini, Chiummo, Cho, Cho, Cho, Christensen, Chu, Chu, Chu, Chua,
  Chung, Chung, Ciani, Ciobanu, Ciolfi, Cipriano, Cirone, Clara, Clark,
  Clearwater, Cleva, Cocchieri, Coccia, Cohadon, Cohen, Colgan, Colleoni,
  Collette, Collins, Cominsky, Constancio, Conti, Cooper, Corban, Corbitt,
  Cordero-Carrión, Corley, Cornish, Corsi, Cortese, Costa, Cotesta, Coughlin,
  Coughlin, Coulon, Countryman, Couvares, Covas, Cowan, Coward, Cowart, Coyne,
  Coyne, Creighton, Creighton, Cripe, Croquette, Crowder, Cullen, Cumming,
  Cunningham, Cuoco, Canton, Dálya, Danilishin, D’Antonio, Danzmann,
  Dasgupta, Da~Silva~Costa, Datrier, Dattilo, Dave, Davier, Davis, Daw, DeBra,
  Deenadayalan, Degallaix, De~Laurentis, Deléglise, Pozzo, DeMarchi, Demos,
  Dent, De~Pietri, Derby, De~Rosa, De~Rossi, DeSalvo, de~Varona, Dhurandhar,
  Díaz, Dietrich, Fiore, Giovanni, Girolamo, Lieto, Ding, Pace, Palma, Renzo,
  Dmitriev, Doctor, Doi, Donovan, Dooley, Doravari, Dorrington, Downes, Drago,
  Driggers, Du, Ducoin, Dupej, Dwyer, Easter, Edo, Edwards, Effler, Eguchi,
  Ehrens, Eichholz, Eikenberry, Eisenmann, Eisenstein, Enomoto, Essick,
  Estelles, Estevez, Etienne, Etzel, Evans, Evans, Fafone, Fair, Fairhurst,
  Fan, Farinon, Farr, Farr, Fauchon-Jones, Favata, Fays, Fazio, Fee, Feicht,
  Fejer, Feng, Fernandez-Galiana, Ferrante, Ferreira, Ferreira, Ferrini,
  Fidecaro, Fiori, Fiorucci, Fishbach, Fisher, Fishner, Fitz-Axen, Flaminio,
  Fletcher, Flynn, Fong, Font, Forsyth, Fournier, Frasca, Frasconi, Frei,
  Freise, Frey, Frey, Fritschel, Frolov, Fujii, Fukunaga, Fukushima, Fulda,
  Fyffe, Gabbard, Gadre, Gaebel, Gair, Gammaitoni, Ganija, Gaonkar, Garcia,
  García-Quirós, Garufi, Gateley, Gaudio, Gaur, Gayathri, Ge, Gemme, Genin,
  Gennai, George, George, Gergely, Germain, Ghonge, Ghosh, Ghosh, Ghosh,
  Giacomazzo, Giaime, Giardina, Giazotto, Gill, Giordano, Glover, Godwin,
  Goetz, Goetz, Goncharov, González, Castro, Gopakumar, Gorodetsky, Gossan,
  Gosselin, Gouaty, Grado, Graef, Granata, Grant, Gras, Grassia, Gray, Gray,
  Greco, Green, Green, Gretarsson, Groot, Grote, Grunewald, Gruning, Guidi,
  Gulati, Guo, Gupta, Gupta, Gustafson, Gustafson, Haegel, Hagiwara, Haino,
  Halim, Hall, Hall, Hamilton, Hammond, Haney, Hanke, Hanks, Hanna, Hannam,
  Hannuksela, Hanson, Hardwick, Haris, Harms, Harry, Harry, Hasegawa, Haster,
  Haughian, Hayakawa, Hayama, Hayes, Healy, Heidmann, Heintze, Heitmann, Hello,
  Hemming, Hendry, Heng, Hennig, Heptonstall, Heurs, Hild, Himemoto, Hinderer,
  Hiranuma, Hirata, Hirose, Hoak, Hochheim, Hofman, Holgado, Holland, Holt,
  Holz, Hong, Hopkins, Horst, Hough, Howell, Hoy, Hreibi, Hsieh, Huang, Huang,
  Huang, Huerta, Huet, Hughey, Hulko, Husa, Huttner, Huynh-Dinh, Idzkowski,
  Iess, Ikenoue, Imam, Inayoshi, Ingram, Inoue, Inta, Intini, Ioka, Irwin, Isa,
  Isac, Isi, Itoh, Iyer, Izumi, Jacqmin, Jadhav, Jani, Janthalur, Jaranowski,
  Jenkins, Jiang, Johnson, Jones, Jones, Jones, Jonker, Ju, Jung, Jung, Junker,
  Kajita, Kalaghatgi, Kalogera, Kamai, Kamiizumi, Kanda, Kandhasamy, Kang,
  Kanner, Kapadia, Karki, Karvinen, Kashyap, Kasprzack, Katsanevas,
  Katsavounidis, Katzman, Kaufer, Kawabe, Kawaguchi, Kawai, Kawasaki,
  Keerthana, Kéfélian, Keitel, Kennedy, Key, Khalili, Khan, Khan, Khan, Khan,
  Khazanov, Khursheed, Kijbunchoo, Kim, Kim, Kim, Kim, Kim, Kim, Kim, Kim,
  Kimball, Kimura, King, King, Kinley-Hanlon, Kirchhoff, Kissel, Kita,
  Kitazawa, Kleybolte, Klika, Klimenko, Knowles, Knyazev, Koch, Koehlenbeck,
  Koekoek, Kojima, Kokeyama, Koley, Komori, Kondrashov, Kong, Kontos, Koper,
  Korobko, Korth, Kotake, Kowalska, Kozak, Kozakai, Kozu, Kringel, Krishnendu,
  Królak, Kuehn, Kumar, Kumar, Kumar, Kumar, Kumar, Kume, Kuo, Kuo, Kuo,
  Kuroyanagi, Kusayanagi, Kutynia, Kwak, Kwang, Lackey, Lai, Lam, Landry, Lane,
  Lang, Lange, Lantz, Lanza, Lartaux-Vollard, Lasky, Laxen, Lazzarini, Lazzaro,
  Leaci, Leavey, Lecoeuche, Lee, Lee, Lee, Lee, Lee, Lee, Lee, Lehmann, Lenon,
  Leonardi, Leroy, Letendre, Levin, Li, Li, Li, Li, Lin, Lin, Lin, Lin, Linde,
  Linker, Littenberg, Liu, Liu, Liu, Lo, Lockerbie, London, Longo, Lorenzini,
  Loriette, Lormand, Losurdo, Lough, Lousto, Lovelace, Lower, Lück, Lumaca,
  Lundgren, Luo, Lynch, Ma, Macas, Macfoy, MacInnis, Macleod, Macquet,
  Magaña-Sandoval, Zertuche, Magee, Majorana, Maksimovic, Malik, Man, Mandic,
  Mangano, Mansell, Manske, Mantovani, Marchesoni, Marchio, Marion, Márka,
  Márka, Markakis, Markosyan, Markowitz, Maros, Marquina, Marsat, Martelli,
  Martin, Martin, Martynov, Mason, Massera, Masserot, Massinger, Masso-Reid,
  Mastrogiovanni, Matas, Matichard, Matone, Mavalvala, Mazumder, McCann,
  McCarthy, McClelland, McCormick, McCuller, McGuire, McIver, McManus, McRae,
  McWilliams, Meacher, Meadors, Mehmet, Mehta, Meidam, Melatos, Mendell,
  Mercer, Mereni, Merilh, Merzougui, Meshkov, Messenger, Messick, Metzdorff,
  Meyers, Miao, Michel, Michimura, Middleton, Mikhailov, Milano, Miller,
  Miller, Millhouse, Mills, Milovich-Goff, Minazzoli, Minenkov, Mio, Mishkin,
  Mishra, Mistry, Mitra, Mitrofanov, Mitselmakher, Mittleman, Miyakawa,
  Miyamoto, Miyazaki, Miyo, Miyoki, Mo, Moffa, Mogushi, Mohapatra, Montani,
  Moore, Moraru, Moreno, Morisaki, Moriwaki, Mours, Mow-Lowry, Mukherjee,
  Mukherjee, Mukherjee, Mukund, Mullavey, Munch, Muñiz, Muratore, Murray,
  Nagano, Nagano, Nagar, Nakamura, Nakano, Nakano, Nakashima, Nardecchia,
  Narikawa, Naticchioni, Nayak, Negishi, Neilson, Nelemans, Nelson, Nery,
  Neunzert, Ng, Ng, Nguyen, Ni, Nichols, Nishizawa, Nissanke, Nocera, North,
  Nuttall, Obergaulinger, Oberling, O’Brien, Obuchi, O’Dea, Ogaki, Ogin,
  Oh, Oh, Ohashi, Ohishi, Ohkawa, Ohme, Ohta, Okada, Okutomi, Oliver, Oohara,
  Ooi, Oppermann, Oram, O’Reilly, Ormiston, Ortega, O’Shaughnessy, Oshino,
  Ossokine, Ottaway, Overmier, Owen, Pace, Pagano, Page, Pai, Pai, Palamos,
  Palashov, Palomba, Pal-Singh, Pan, Pan, Pang, Pang, Pang, Pankow, Pannarale,
  Pant, Paoletti, Paoli, Papa, Parida, Park, Parker, Pascucci, Pasqualetti,
  Passaquieti, Passuello, Patil, Patricelli, Pearlstone, Pedersen, Pedraza,
  Pedurand, Pele, Arellano, Penn, Perez, Perreca, Pfeiffer, Phelps, Phukon,
  Piccinni, Pichot, Piergiovanni, Pillant, Pinard, Pinto, Pirello, Pitkin,
  Poggiani, Pong, Ponrathnam, Popolizio, Porter, Powell, Prajapati, Prasad,
  Prasai, Prasanna, Pratten, Prestegard, Privitera, Prodi, Prokhorov, Puncken,
  Punturo, Puppo, Pürrer, Qi, Quetschke, Quinonez, Quintero, Quitzow-James,
  Raab, Radkins, Radulescu, Raffai, Raja, Rajan, Rajbhandari, Rakhmanov,
  Ramirez, Ramos-Buades, Rana, Rao, Rapagnani, Raymond, Razzano, Read,
  Regimbau, Rei, Reid, Reitze, Ren, Ricci, Richardson, Richardson, Ricker,
  Riles, Rizzo, Robertson, Robie, Robinet, Rocchi, Rolland, Rollins, Roma,
  Romanelli, Romano, Romel, Romie, Rose, Rosińska, Rosofsky, Ross, Rowan,
  Rüdiger, Ruggi, Rutins, Ryan, Sachdev, Sadecki, Sago, Saito, Saito, Sakai,
  Sakai, Sakamoto, Sakellariadou, Sakuno, Salconi, Saleem, Samajdar, Sammut,
  Sanchez, Sanchez, Sanchis-Gual, Sandberg, Sanders, Santiago, Sarin, Sassolas,
  Sathyaprakash, Sato, Sato, Sauter, Savage, Sawada, Schale, Scheel, Scheuer,
  Schmidt, Schnabel, Schofield, Schönbeck, Schreiber, Schulte, Schutz,
  Schwalbe, Scott, Scott, Seidel, Sekiguchi, Sekiguchi, Sellers, Sengupta,
  Sennett, Sentenac, Sequino, Sergeev, Setyawati, Shaddock, Shaffer, Shahriar,
  Shaner, Shao, Sharma, Shawhan, Shen, Shibagaki, Shimizu, Shimoda, Shimode,
  Shink, Shinkai, Shishido, Shoda, Shoemaker, Shoemaker, ShyamSundar, Siellez,
  Sieniawska, Sigg, Silva, Singer, Singh, Singhal, Sintes, Sitmukhambetov,
  Skliris, Slagmolen, Slaven-Blair, Smith, Smith, Somala, Somiya, Son, Sorazu,
  Sorrentino, Sotani, Souradeep, Sowell, Spencer, Srivastava, Srivastava,
  Staats, Stachie, Standke, Steer, Steinke, Steinlechner, Steinlechner,
  Steinmeyer, Stevenson, Stocks, Stone, Stops, Strain, Stratta, Strigin,
  Strunk, Sturani, Stuver, Sudhir, Sugimoto, Summerscales, Sun, Sunil, Suresh,
  Sutton, Suzuki, Suzuki, Swinkels, Szczepańczyk, Tacca, Tagoshi, Tait,
  Takahashi, Takahashi, Takamori, Takano, Takeda, Takeda, Talbot, Talukder,
  Tanaka, Tanaka, Tanaka, Tanaka, Tanaka, Tanioka, Tanner, Tápai, Martin,
  Taracchini, Tasson, Taylor, Telada, Thies, Thomas, Thomas, Thondapu, Thorne,
  Thrane, Tiwari, Tiwari, Tiwari, Toland, Tomaru, Tomigami, Tomura, Tonelli,
  Tornasi, Torres-Forné, Torrie, Töyrä, Travasso, Traylor, Tringali,
  Trovato, Trozzo, Trudeau, Tsang, Tsang, Tse, Tso, Tsubono, Tsuchida, Tsukada,
  Tsuna, Tsuzuki, Tuyenbayev, Uchikata, Uchiyama, Ueda, Uehara, Ueno, Ueshima,
  Ugolini, Unnikrishnan, Uraguchi, Urban, Ushiba, Usman, Vahlbruch, Vajente,
  Valdes, Bakel, Beuzekom, Brand, Broeck, Vander-Hyde, Schaaf, Heijningen,
  Putten, Veggel, Vardaro, Varma, Vass, Vasúth, Vecchio, Vedovato, Veitch,
  Veitch, Venkateswara, Venugopalan, Verkindt, Vetrano, Viceré, Viets, Vine,
  Vinet, Vitale, Vivanco, Vo, Vocca, Vorvick, Vyatchanin, Wade, Wade, Wade,
  Walet, Walker, Wallace, Walsh, Wang, Wang, Wang, Wang, Wang, Wang, Ward,
  Warden, Warner, Was, Watchi, Weaver, Wei, Weinert, Weinstein, Weiss,
  Wellmann, Wen, Wessel, Weßels, Westhouse, Wette, Whelan, Whiting, Whittle,
  Wilken, Williams, Williamson, Willis, Willke, Wimmer, Winkler, Wipf, Wittel,
  Woan, Woehler, Wofford, Worden, Wright, Wu, Wu, Wu, Wu, Wysocki, Xiao, Xu,
  Yamada, Yamamoto, Yamamoto, Yamamoto, Yamamoto, Yancey, Yang, Yap, Yazback,
  Yeeles, Yokogawa, Yokoyama, Yokozawa, Yoshioka, Yu, Yu, Yuen, Yuzurihara,
  Yvert, Zadrożny, Zanolin, Zeidler, Zelenova, Zendri, Zevin, Zhang, Zhang,
  Zhang, Zhao, Zhao, Zhou, Zhou, Zhu, Zhu, Zimmerman, Zucker, \&
  Zweizig}]{detector_observing_prospects}
---. 2020, Living Reviews in Relativity, 23, \dodoi{10.1007/s41114-020-00026-9}

\bibitem[{Abbott {et~al.}(2021{\natexlab{a}})Abbott, Abe, Acernese, Ackley,
  Adhikari, Adhikari, Adkins, Adya, Affeldt, Agarwal, Agathos, Agatsuma,
  Aggarwal, Aguiar, Aiello, Ain, Ajith, Akutsu, de~Alarcón, Albanesi, Alfaidi,
  Allocca, Altin, Amato, Anand, Anand, Ananyeva, Anderson, Anderson, Ando,
  Andrade, Andres, Andrés-Carcasona, Andrić, Angelova, Ansoldi, Antelis,
  Antier, Apostolatos, Appavuravther, Appert, Apple, Arai, Araya, Araya,
  Areeda, Arène, Aritomi, Arnaud, Arogeti, Aronson, Arun, Asada, Asali,
  Ashton, Aso, Assiduo, de~Souza~Melo, Aston, Astone, Aubin, AultONeal, Austin,
  Babak, Badaracco, Bader, Badger, Bae, Bae, Baer, Bagnasco, Bai, Baird,
  Bajpai, Baka, Ball, Ballardin, Ballmer, Balsamo, Baltus, Banagiri, Banerjee,
  Bankar, Barayoga, Barbieri, Barish, Barker, Barneo, Barone, Barr, Barsotti,
  Barsuglia, Barta, Bartlett, Barton, Bartos, Basak, Bassiri, Basti, Bawaj,
  Bayley, Bazzan, Becher, Bécsy, Bedakihale, Beirnaert, Bejger, Belahcene,
  Benedetto, Beniwal, Benjamin, Bennett, Bentley, BenYaala, Bera, Berbel,
  Bergamin, Berger, Bernuzzi, Berry, Bersanetti, Bertolini, Betzwieser,
  Beveridge, Bhandare, Bhandari, Bhardwaj, Bhatt, Bhattacharjee, Bhaumik,
  Bianchi, Bilenko, Billingsley, Bini, Birney, Birnholtz, Biscans, Bischi,
  Biscoveanu, Bisht, Biswas, Bitossi, Bizouard, Blackburn, Blair, Blair, Blair,
  Bobba, Bode, Boër, Bogaert, Boldrini, Bolingbroke, Bonavena, Bondu, Bonilla,
  Bonnand, Booker, Boom, Bork, Boschi, Bose, Bose, Bossilkov, Boudart,
  Bouffanais, Bozzi, Bradaschia, Brady, Bramley, Branch, Branchesi, Brau,
  Breschi, Briant, Briggs, Brillet, Brinkmann, Brockill, Brooks, Brooks, Brown,
  Brunett, Bruno, Bruntz, Bryant, Bucci, Bulik, Bulten, Buonanno, Burtnyk,
  Buscicchio, Buskulic, Buy, Byer, Davies, Cabras, Cabrita, Cadonati, Caesar,
  Cagnoli, Cahillane, Bustillo, Callaghan, Callister, Calloni, Cameron, Camp,
  Canepa, Canevarolo, Cannavacciuolo, Cannon, Cao, Cao, Capocasa, Capote,
  Carapella, Carbognani, Carlassara, Carlin, Carney, Carpinelli, Carrillo,
  Carullo, Carver, Diaz, Casentini, Castaldi, Caudill, Cavaglià, Cavalier,
  Cavalieri, Cella, Cerdá-Durán, Cesarini, Chaibi, Subrahmanya, Champion,
  Chan, Chan, Chan, Chan, Chan, Chandra, Chang, Chanial, Chao, Chapman-Bird,
  Charlton, Chase, Chassande-Mottin, Chatterjee, Chatterjee, Chatterjee,
  Chaturvedi, Chaty, Chatziioannou, Chen, Chen, Chen, Chen, Chen, Chen, Chen,
  Chen, Chen, Cheng, Cheong, Cheung, Chia, Chiadini, Chiang, Chiarini,
  Chierici, Chincarini, Chiofalo, Chiummo, Choudhary, Choudhary, Christensen,
  Chu, Chu, Chua, Chung, Ciani, Ciecielag, Cieślar, Cifaldi, Ciobanu, Ciolfi,
  Cipriano, Clara, Clark, Clearwater, Clesse, Cleva, Coccia, Codazzo, Cohadon,
  Cohen, Colleoni, Collette, Colombo, Colpi, Compton, au2, Conti, Cooper,
  Corban, Corbitt, Cordero-Carrión, Corezzi, Corley, Cornish, Corre, Corsi,
  Cortese, Costa, Cotesta, Cottingham, Coughlin, Coulon, Countryman, Cousins,
  Couvares, Coward, Cowart, Coyne, Coyne, Creighton, Creighton, Criswell,
  Croquette, Crowder, Cudell, Cullen, Cumming, Cummings, Cunningham, Cuoco,
  Curyło, Dabadie, Canton, Dall'Osso, Dálya, Dana, D'Angelo, Danilishin,
  D'Antonio, Danzmann, Darsow-Fromm, Dasgupta, Datrier, Datta, Datta, Dattilo,
  Dave, Davier, Davis, Davis, Daw, Dean, DeBra, Deenadayalan, Degallaix,
  Laurentis, Deléglise, Favero, Lillo, Lillo, Dell'Aquila, Pozzo, DeMarchi,
  Matteis, D'Emilio, Demos, Dent, Depasse, Pietri, Rosa, Rossi, DeSalvo,
  Simone, Dhurandhar, Díaz, Didio, Dietrich, Fiore, Fronzo, Giorgio, Giovanni,
  Giovanni, Girolamo, Lieto, Michele, Ding, Pace, Palma, Renzo, Divakarla,
  Divyajyoti, Dmitriev, Doctor, Donahue, D'Onofrio, Donovan, Dooley, Doravari,
  Drago, Driggers, Drori, Ducoin, Dupej, Dupletsa, Durante, D'Urso, Duverne,
  Dwyer, Eassa, Easter, Ebersold, Eckhardt, Eddolls, Edelman, Edo, Edy, Effler,
  Eguchi, Eichholz, Eikenberry, Eisenmann, Eisenstein, Ejlli, Engelby, Enomoto,
  Errico, Essick, Estellés, Estevez, Etienne, Etzel, Evans, Evans, Evstafyeva,
  Ewing, Fabrizi, Faedi, Fafone, Fair, Fairhurst, Fan, Farah, Farinon, Farr,
  Farr, Fauchon-Jones, Favaro, Favata, Fays, Fazio, Feicht, Fejer, Fenyvesi,
  Ferguson, Fernandez-Galiana, Ferrante, Ferreira, Fidecaro, Figura, Fiori,
  Fiori, Fishbach, Fisher, Fittipaldi, Fiumara, Flaminio, Floden, Fong, Font,
  Fornal, Forsyth, Franke, Frasca, Frasconi, Freed, Frei, Freise, Freitas,
  Frey, Fritschel, Frolov, Fronzé, Fujii, Fujikawa, Fujimoto, Fulda, Fyffe,
  Gabbard, Gabella, Gadre, Gair, Gais, Galaudage, Gamba, Ganapathy, Ganguly,
  Gao, Gaonkar, Garaventa, Núñez, García-Quirós, Garufi, Gateley, Gayathri,
  Ge, Gemme, Gennai, George, Gerberding, Gergely, Gewecke, Ghonge, Ghosh,
  Ghosh, Ghosh, Ghosh, Ghosh, Giacomazzo, Giacoppo, Giaime, Giardina, Gibson,
  Gier, Giesler, Giri, Gissi, Gkaitatzis, Glanzer, Gleckl, Godwin, Goetz,
  Goetz, Gohlke, Golomb, Goncharov, González, Gosselin, Gouaty, Gould, Goyal,
  Grace, Grado, Graham, Granata, Granata, Grant, Gras, Grassia, Gray, Gray,
  Greco, Green, Green, Gretarsson, Gretarsson, Griffith, Griffiths, Griggs,
  Grignani, Grimaldi, Grimes, Grimm, Grote, Grunewald, Gruning, Gruson, Guerra,
  Guidi, Guimaraes, Guixé, Gulati, Gunny, Guo, Guo, Gupta, Gupta, Gupta,
  Gupta, Gupta, Gustafson, Guzman, Ha, Hadiputrawan, Haegel, Haino, Halim,
  Hall, Hamilton, Hammond, Han, Haney, Hanks, Hanna, Hannam, Hannuksela,
  Hansen, Hansen, Hanson, Harder, Haris, Harms, Harry, Harry, Hartwig,
  Hasegawa, Haskell, Haster, Hathaway, Hattori, Haughian, Hayakawa, Hayama,
  Hayes, Healy, Heidmann, Heidt, Heintze, Heinze, Heinzel, Heitmann, Hellman,
  Hello, Helmling-Cornell, Hemming, Hendry, Heng, Hennes, Hennig, Hennig,
  Henshaw, Hernandez, Vivanco, Heurs, Hewitt, Higginbotham, Hild, Hill,
  Himemoto, Hines, Hirata, Hirose, Ho, Hochheim, Hofman, Hohmann, Holcomb,
  Holland, Hollows, Holmes, Holt, Holz, Hong, Hough, Hourihane, Howell, Hoy,
  Hoyland, Hreibi, Hsieh, Hsieh, Hsiung, Hsu, Huang, Huang, Huang, Huang,
  Huang, Huang, Hübner, Huddart, Hughey, Hui, Hui, Husa, Huttner, Huxford,
  Huynh-Dinh, Ide, Idzkowski, Iess, Inayoshi, Inoue, Iosif, Isi, Isleif, Ito,
  Itoh, Iyer, JaberianHamedan, Jacqmin, Jacquet, Jadhav, Jadhav, Jain, James,
  Jan, Jani, Janquart, Janssens, Janthalur, Jaranowski, Jariwala, Jaume,
  Jenkins, Jenner, Jeon, Jia, Jiang, Jin, Johns, Johnson-McDaniel, Johnston,
  Jones, Jones, Jones, Jones, Joshi, Ju, Jue, Jung, Jung, Junker, Juste,
  Kaihotsu, Kajita, Kakizaki, Kalaghatgi, Kalogera, Kamai, Kamiizumi, Kanda,
  Kandhasamy, Kang, Kanner, Kao, Kapadia, Kapasi, Karathanasis, Karki, Kashyap,
  Kasprzack, Kastaun, Kato, Katsanevas, Katsavounidis, Katzman, Kaur, Kawabe,
  Kawaguchi, Kéfélian, Keitel, Key, Khadka, Khalili, Khan, Khanam, Khazanov,
  Khetan, Khursheed, Kijbunchoo, Kim, Kim, Kim, Kim, Kim, Kim, Kim, Kimball,
  Kimura, Kinley-Hanlon, Kirchhoff, Kissel, Klimenko, Klinger, Knee, Knowles,
  Knust, Knyazev, Kobayashi, Koch, Koekoek, Kohri, Kokeyama, Koley, Kolitsidou,
  Kolstein, Komori, Kondrashov, Kong, Kontos, Koper, Korobko, Kovalam, Koyama,
  Kozak, Kozakai, Kringel, Krishnendu, Królak, Kuehn, Kuei, Kuijer, Kulkarni,
  Kumar, Kumar, Kumar, Kumar, Kume, Kuns, Kuromiya, Kuroyanagi, Kwak, Lacaille,
  Lagabbe, Laghi, Lalande, Lalleman, Lam, Lamberts, Landry, Lane, Lang, Lange,
  Lantz, Rosa, Lartaux-Vollard, Lasky, Laxen, Lazzarini, Lazzaro, Leaci,
  Leavey, LeBohec, Lecoeuche, Lee, Lee, Lee, Lee, Lee, Legred, Lehmann,
  Lemaître, Lenti, Leonardi, Leonova, Leroy, Letendre, Levesque, Levin,
  Leviton, Leyde, Li, Li, Li, Li, Li, Li, Li, Lin, Lin, Lin, Lin, Lin, Lin,
  Linde, Linker, Linley, Littenberg, Liu, Liu, Liu, Liu, Llamas, Lo, Lo,
  London, Longo, Lopez, Portilla, Lorenzini, Loriette, Lormand, Losurdo, Lott,
  Lough, Lousto, Lovelace, Lucaccioni, Lück, Lumaca, Lundgren, Luo, Lynam,
  Ma'arif, Macas, Machtinger, MacInnis, Macleod, MacMillan, Macquet, Hernandez,
  Magazzù, Magee, Maggiore, Magnozzi, Mahesh, Majorana, Maksimovic, Maliakal,
  Malik, Man, Mandic, Mangano, Mansell, Manske, Mantovani, Mapelli, Marchesoni,
  Pina, Marion, Mark, Márka, Márka, Markakis, Markosyan, Markowitz, Maros,
  Marquina, Marsat, Martelli, Martin, Martin, Martinez, Martinez, Martinez,
  Martinovic, Martynov, Marx, Masalehdan, Mason, Massera, Masserot, Masso-Reid,
  Mastrogiovanni, Matas, Mateu-Lucena, Matichard, Matiushechkina, Mavalvala,
  McCann, McCarthy, McClelland, McClincy, McCormick, McCuller, McGhee, McGuire,
  McIsaac, McIver, McRae, McWilliams, Meacher, Mehmet, Mehta, Meijer, Melatos,
  Melchor, Mendell, Menendez-Vazquez, Menoni, Mercer, Mereni, Merfeld, Merilh,
  Merritt, Merzougui, Meshkov, Messenger, Messick, Meyers, Meylahn, Mhaske,
  Miani, Miao, Michaloliakos, Michel, Michimura, Middleton, Mihaylov, Milano,
  Miller, Miller, Miller, Millhouse, Mills, Milotti, Minenkov, Mio, Mir,
  Miravet-Tenés, Mishkin, Mishra, Mishra, Mistry, Mitra, Mitrofanov,
  Mitselmakher, Mittleman, Miyakawa, Miyo, Miyoki, Mo, Modafferi, Moguel,
  Mogushi, Mohapatra, Mohite, Molina, Molina-Ruiz, Mondin, Montani, Moore,
  Moragues, Moraru, Morawski, More, Moreno, Moreno, Mori, Morisaki, Morisue,
  Moriwaki, Mours, Mow-Lowry, Mozzon, Muciaccia, Mukherjee, Mukherjee,
  Mukherjee, Mukherjee, Mukherjee, Mukund, Mullavey, Munch, Muñiz, Murray,
  Musenich, Muusse, Nadji, Nagano, Nagar, Nakamura, Nakano, Nakano, Nakayama,
  Napolano, Nardecchia, Narikawa, Narola, Naticchioni, Nayak, Nayak, Neil,
  Neilson, Nelson, Nelson, Nery, Neubauer, Neunzert, Ng, Ng, Nguyen, Nguyen,
  Nguyen, Quynh, Ni, Ni, Nichols, Nishimoto, Nishizawa, Nissanke, Nitoglia,
  Nocera, Norman, North, Nozaki, Nurbek, Nuttall, Obayashi, Oberling, O'Brien,
  O'Dell, Oelker, Ogaki, Oganesyan, Oh, Oh, Oh, Ohashi, Ohashi, Ohkawa, Ohme,
  Ohta, Okada, Okutani, Olivetto, Oohara, Oram, O'Reilly, Ormiston, Ormsby,
  O'Shaughnessy, O'Shea, Oshino, Ossokine, Osthelder, Otabe, Ottaway, Overmier,
  Pace, Pagano, Pagano, Page, Pagliaroli, Pai, Pai, Pal, Palamos, Palashov,
  Palomba, Pan, Pan, Panda, Pang, Pankow, Pannarale, Pant, Panther, Paoletti,
  Paoli, Paolone, Pappas, Parisi, Park, Park, Parker, Pascucci, Pasqualetti,
  Passaquieti, Passuello, Patel, Pathak, Patricelli, Patron, Paul, Payne,
  Pedraza, Pedurand, Pegoraro, Pele, Arellano, Penano, Penn, Perego, Pereira,
  Pereira, Perez, Périgois, Perkins, Perreca, Perriès, Pesios, Petermann,
  Petterson, Pfeiffer, Pham, Pham, Phukon, Phurailatpam, Piccinni, Pichot,
  Piendibene, Piergiovanni, Pierini, Pierro, Pillant, Pillas, Pilo, Pinard,
  Pineda-Bosque, Pinto, Pinto, Piotrzkowski, Piotrzkowski, Pirello, Pitkin,
  Placidi, Placidi, Planas, Plastino, Pluchar, Poggiani, Polini, Pong,
  Ponrathnam, Porter, Poulton, Poverman, Powell, Pracchia, Pradier, Prajapati,
  Prasai, Prasanna, Pratten, Principe, Prodi, Prokhorov, Prosposito, Prudenzi,
  Puecher, Punturo, Puosi, Puppo, Pürrer, Qi, Quartey, Quetschke, Quinonez,
  Quitzow-James, Qutob, Raab, Raaijmakers, Radkins, Radulesco, Raffai, Rail,
  Raja, Rajan, Ramirez, Ramirez, Ramos-Buades, Rana, Rapagnani, Ray, Raymond,
  Raza, Razzano, Read, Rees, Regimbau, Rei, Reid, Reid, Reitze, Relton,
  Renzini, Rettegno, Revenu, Reza, Rezac, Ricci, Richards, Richardson,
  Richardson, Riemenschneider, Riles, Rinaldi, Rink, Robertson, Robie, Robinet,
  Rocchi, Rodriguez, Rolland, Rollins, Romanelli, Romano, Romel, Romero,
  Romero-Shaw, Romie, Ronchini, Rosa, Rose, Rosińska, Ross, Rowan, Rowlinson,
  Roy, Roy, Roy, Rozza, Ruggi, Ruiz-Rocha, Ryan, Sachdev, Sadecki, Sadiq, Saha,
  Saito, Sakai, Sakellariadou, Sakon, Salafia, Salces-Carcoba, Salconi, Saleem,
  Salemi, Samajdar, Sanchez, Sanchez, Sanchez, Sanchis-Gual, Sanders, Sanuy,
  Saravanan, Sarin, Sassolas, Satari, Sathyaprakash, Sauter, Savage, Savant,
  Sawada, Sawant, Sayah, Schaetzl, Scheel, Scheuer, Schiworski, Schmidt,
  Schmidt, Schnabel, Schneewind, Schofield, Schönbeck, Schulte, Schutz,
  Schwartz, Scott, Scott, Seglar-Arroyo, Sekiguchi, Sellers, Sengupta,
  Sentenac, Seo, Sequino, Sergeev, Setyawati, Shaffer, Shahriar, Shaikh, Shams,
  Shao, Sharma, Sharma, Shawhan, Shcheblanov, Sheela, Shikano, Shikauchi,
  Shimizu, Shimode, Shinkai, Shishido, Shoda, Shoemaker, Shoemaker,
  ShyamSundar, Sieniawska, Sigg, Silenzi, Singer, Singh, Singh, Singh, Singha,
  Sintes, Sipala, Skliris, Slagmolen, Slaven-Blair, Smetana, Smith, Smith,
  Smith, Soldateschi, Somala, Somiya, Song, Soni, Soni, Sordini, Sorrentino,
  Sorrentino, Soulard, Souradeep, Sowell, Spagnuolo, Spencer, Spera,
  Spinicelli, Srivastava, Srivastava, Staats, Stachie, Stachurski, Steer,
  Steinhoff, Steinlechner, Steinlechner, Stergioulas, Stops, Stover, Strain,
  Strang, Stratta, Strong, Strunk, Sturani, Stuver, Suchenek, Sudhagar, Sudhir,
  Sugimoto, Suh, Sullivan, Sullivan, Summerscales, Sun, Sunil, Sur, Suresh,
  Sutton, Suzuki, Suzuki, Suzuki, Swinkels, Szczepańczyk, Szewczyk, Tacca,
  Tagoshi, Tait, Takahashi, Takahashi, Takano, Takeda, Takeda, Talbot, Talbot,
  Tanaka, Tanaka, Tanaka, Tanasijczuk, Tanioka, Tanner, Tao, Tao, Tapia,
  Martín, Taranto, Taruya, Tasson, Tenorio, Terhune, Terkowski,
  Thirugnanasambandam, Thomas, Thomas, Thompson, Thompson, Thondapu, Thorne,
  Thrane, Tiwari, Tiwari, Tiwari, Toivonen, Tolley, Tomaru, Tomura, Tonelli,
  Tornasi, Torres-Forné, Torrie, e~Melo, Töyrä, Trapananti, Travasso,
  Traylor, Trevor, Tringali, Tripathee, Troiano, Trovato, Trozzo, Trudeau,
  Tsai, Tsang, Tsang, Tsao, Tse, Tso, Tsuchida, Tsukada, Tsuna, Tsutsui,
  Turbang, Turconi, Tuyenbayev, Ubhi, Uchikata, Uchiyama, Udall, Ueda, Uehara,
  Ueno, Ueshima, Unnikrishnan, Urban, Ushiba, Utina, Vajente, Vajpeyi, Valdes,
  Valentini, Valsan, van Bakel, van Beuzekom, van Dael, van~den Brand, Broeck,
  Vander-Hyde, van Haevermaet, van Heijningen, van Putten, van Remortel,
  Vardaro, Vargas, Varma, Vasúth, Vecchio, Vedovato, Veitch, Veitch,
  Venneberg, Venugopalan, Verkindt, Verma, Verma, Vermeulen, Veske, Vetrano,
  Viceré, Vidyant, Viets, Vijaykumar, Villa-Ortega, Vinet, Virtuoso, Vitale,
  Vocca, von Reis, von Wrangel, Vorvick, Vyatchanin, Wade, Wade, Wagner, Wald,
  Walet, Walker, Wallace, Wallace, Wang, Wang, Wang, Ward, Warner, Was,
  Washimi, Washington, Watchi, Weaver, Weaving, Webster, Weinert, Weinstein,
  Weiss, Weller, Weller, Wellmann, Wen, Weßels, Wette, Whelan, White, Whiting,
  Whittle, Wilken, Williams, Williams, Williamson, Willis, Willke, Wilson,
  Wipf, Wlodarczyk, Woan, Woehler, Wofford, Wong, Wong, Wright, Wu, Wu, Wu,
  Wysocki, Xiao, Yamada, Yamamoto, Yamamoto, Yamamoto, Yamashita, Yamazaki,
  Yang, Yang, Yang, Yang, Yang, Yang, Yap, Yeeles, Yeh, Yelikar, Ying,
  Yokoyama, Yokozawa, Yoo, Yoshioka, Yu, Yu, Yuzurihara, Zadrożny, Zanolin,
  Zeidler, Zelenova, Zendri, Zevin, Zhan, Zhang, Zhang, Zhang, Zhang, Zhang,
  Zhang, Zhao, Zhao, Zhao, Zhao, Zhou, Zhou, Zhu, Zhu, Zimmerman, Zucker, \&
  Zweizig}]{o3_tgr}
Abbott, R., Abe, H., Acernese, F., {et~al.} 2021{\natexlab{a}}, Tests of
  General Relativity with GWTC-3.
\newblock \doarXiv{2112.06861}

\bibitem[{Abbott {et~al.}(2021{\natexlab{b}})}]{LIGOScientific:2021izm}
Abbott, R., {et~al.} 2021{\natexlab{b}}, Astrophys. J., 923, 14,
  \dodoi{10.3847/1538-4357/ac23db}

\bibitem[{Abbott {et~al.}(2023{\natexlab{a}})Abbott, Abbott, Acernese, Ackley,
  Adams, Adhikari, Adhikari, Adya, Affeldt, Agarwal, Agathos, Agatsuma,
  Aggarwal, Aguiar, Aiello, Ain, Ajith, Akcay, Akutsu, Albanesi, Allocca,
  Altin, Amato, Anand, Anand, Ananyeva, Anderson, Anderson, Ando, Andrade,
  Andres, Andrić, Angelova, Ansoldi, Antelis, Antier, Appert, Arai, Arai,
  Arai, Araki, Araya, Araya, Areeda, Arène, Aritomi, Arnaud, Arogeti, Aronson,
  Arun, Asada, Asali, Ashton, Aso, Assiduo, Aston, Astone, Aubin, Austin,
  Babak, Badaracco, Bader, Badger, Bae, Bae, Baer, Bagnasco, Bai, Baiotti,
  Baird, Bajpai, Ball, Ballardin, Ballmer, Balsamo, Baltus, Banagiri, Bankar,
  Barayoga, Barbieri, Barish, Barker, Barneo, Barone, Barr, Barsotti,
  Barsuglia, Barta, Bartlett, Barton, Bartos, Bassiri, Basti, Bawaj, Bayley,
  Baylor, Bazzan, Bécsy, Bedakihale, Bejger, Belahcene, Benedetto, Beniwal,
  Bennett, Bentley, BenYaala, Bergamin, Berger, Bernuzzi, Berry, Bersanetti,
  Bertolini, Betzwieser, Beveridge, Bhandare, Bhardwaj, Bhattacharjee, Bhaumik,
  Bilenko, Billingsley, Bini, Birney, Birnholtz, Biscans, Bischi, Biscoveanu,
  Bisht, Biswas, Bitossi, Bizouard, Blackburn, Blair, Blair, Blair, Bobba,
  Bode, Boer, Bogaert, Boldrini, Bonavena, Bondu, Bonilla, Bonnand, Booker,
  Boom, Bork, Boschi, Bose, Bose, Bossilkov, Boudart, Bouffanais, Bozzi,
  Bradaschia, Brady, Bramley, Branch, Branchesi, Brandt, Brau, Breschi, Briant,
  Briggs, Brillet, Brinkmann, Brockill, Brooks, Brooks, Brown, Brunett, Bruno,
  Bruntz, Bryant, Bulik, Bulten, Buonanno, Buscicchio, Buskulic, Buy, Byer,
  Davies, Cadonati, Cagnoli, Cahillane, Bustillo, Callaghan, Callister,
  Calloni, Cameron, Camp, Canepa, Canevarolo, Cannavacciuolo, Cannon, Cao, Cao,
  Capocasa, Capote, Carapella, Carbognani, Carlin, Carney, Carpinelli,
  Carrillo, Carullo, Carver, Diaz, Casentini, Castaldi, Caudill, Cavaglià,
  Cavalier, Cavalieri, Ceasar, Cella, Cerdá-Durán, Cesarini, Chaibi,
  Chakravarti, Subrahmanya, Champion, Chan, Chan, Chan, Chan, Chan, Chandra,
  Chanial, Chao, Chapman-Bird, Charlton, Chase, Chassande-Mottin, Chatterjee,
  Chatterjee, Chatterjee, Chaturvedi, Chaty, Chatziioannou, Chen, Chen, Chen,
  Chen, Chen, Chen, Chen, Chen, Cheng, Cheong, Cheung, Chia, Chiadini, Chiang,
  Chiarini, Chierici, Chincarini, Chiofalo, Chiummo, Cho, Cho, Choudhary,
  Choudhary, Christensen, Chu, Chu, Chu, Chua, Chung, Ciani, Ciecielag,
  Cieślar, Cifaldi, Ciobanu, Ciolfi, Cipriano, Cirone, Clara, Clark, Clark,
  Clarke, Clearwater, Clesse, Cleva, Coccia, Codazzo, Cohadon, Cohen, Cohen,
  Colleoni, Collette, Colombo, Colpi, Compton, Constancio, Conti, Cooper,
  Corban, Corbitt, Cordero-Carrión, Corezzi, Corley, Cornish, Corre, Corsi,
  Cortese, Costa, Cotesta, Coughlin, Coulon, Countryman, Cousins, Couvares,
  Coward, Cowart, Coyne, Coyne, Creighton, Creighton, Criswell, Croquette,
  Crowder, Cudell, Cullen, Cumming, Cummings, Cunningham, Cuoco, Curyło,
  Dabadie, Canton, Dall’Osso, Dálya, Dana, DaneshgaranBajastani, D’Angelo,
  Danila, Danilishin, D’Antonio, Danzmann, Darsow-Fromm, Dasgupta, Datrier,
  Dattilo, Dave, Davier, Davis, Davis, Daw, de~Alarcón, Dean, DeBra,
  Deenadayalan, Degallaix, De~Laurentis, Deléglise, Del~Favero, De~Lillo,
  De~Lillo, Del~Pozzo, DeMarchi, De~Matteis, D’Emilio, Demos, Dent, Depasse,
  De~Pietri, De~Rosa, De~Rossi, DeSalvo, De~Simone, Dhurandhar, Díaz,
  Diaz-Ortiz, Didio, Dietrich, Di~Fiore, Di~Fronzo, Di~Giorgio, Di~Giovanni,
  Di~Giovanni, Di~Girolamo, Di~Lieto, Ding, Di~Pace, Di~Palma, Di~Renzo,
  Divakarla, Dmitriev, Doctor, D’Onofrio, Donovan, Dooley, Doravari,
  Dorrington, Drago, Driggers, Drori, Ducoin, Dupej, Durante, D’Urso,
  Duverne, Dwyer, Eassa, Easter, Ebersold, Eckhardt, Eddolls, Edelman, Edo,
  Edy, Effler, Eguchi, Eichholz, Eikenberry, Eisenmann, Eisenstein, Ejlli,
  Engelby, Enomoto, Errico, Essick, Estellés, Estevez, Etienne, Etzel, Evans,
  Evans, Ewing, Fafone, Fair, Fairhurst, Farah, Farinon, Farr, Farr, Farrow,
  Fauchon-Jones, Favaro, Favata, Fays, Fazio, Feicht, Fejer, Fenyvesi,
  Ferguson, Fernandez-Galiana, Ferrante, Ferreira, Fidecaro, Figura, Fiori,
  Fishbach, Fisher, Fittipaldi, Fiumara, Flaminio, Floden, Fong, Font, Fornal,
  Forsyth, Franke, Frasca, Frasconi, Frederick, Freed, Frei, Freise, Frey,
  Fritschel, Frolov, Fronzé, Fujii, Fujikawa, Fukunaga, Fukushima, Fulda,
  Fyffe, Gabbard, Gabella, Gadre, Gair, Gais, Galaudage, Gamba, Ganapathy,
  Ganguly, Gao, Gaonkar, Garaventa, García, García-Núñez, García-Quirós,
  Garufi, Gateley, Gaudio, Gayathri, Ge, Gemme, Gennai, George, George,
  Gerberding, Gergely, Gewecke, Ghonge, Ghosh, Ghosh, Ghosh, Ghosh, Giacomazzo,
  Giacoppo, Giaime, Giardina, Gibson, Gier, Giesler, Giri, Gissi, Glanzer,
  Gleckl, Godwin, Goetz, Goetz, Gohlke, Golomb, Goncharov, González,
  Gopakumar, Gosselin, Gouaty, Gould, Grace, Grado, Granata, Granata, Grant,
  Gras, Grassia, Gray, Gray, Greco, Green, Green, Gretarsson, Gretarsson,
  Griffith, Griffiths, Griggs, Grignani, Grimaldi, Grimm, Grote, Grunewald,
  Gruning, Guerra, Guidi, Guimaraes, Guixé, Gulati, Guo, Guo, Gupta, Gupta,
  Gupta, Gustafson, Gustafson, Guzman, Ha, Haegel, Hagiwara, Haino, Halim,
  Hall, Hamilton, Hammond, Han, Haney, Hanks, Hanna, Hannam, Hannuksela,
  Hansen, Hansen, Hanson, Harder, Hardwick, Haris, Harms, Harry, Harry,
  Hartwig, Hasegawa, Haskell, Hasskew, Haster, Hattori, Haughian, Hayakawa,
  Hayama, Hayes, Healy, Heidmann, Heidt, Heintze, Heinze, Heinzel, Heitmann,
  Hellman, Hello, Helmling-Cornell, Hemming, Hendry, Heng, Hennes, Hennig,
  Hennig, Hernandez, Hernandez~Vivanco, Heurs, Hild, Hill, Himemoto, Hines,
  Hiranuma, Hirata, Hirose, Hochheim, Hofman, Hohmann, Holcomb, Holland,
  Holley-Bockelmann, Hollows, Holmes, Holt, Holz, Hong, Hopkins, Hough,
  Hourihane, Howell, Hoy, Hoyland, Hreibi, Hsieh, Hsu, Huang, Huang, Huang,
  Huang, Huang, Huang, Hübner, Huddart, Hughey, Hui, Hui, Husa, Huttner,
  Huxford, Huynh-Dinh, Ide, Idzkowski, Iess, Ikenoue, Imam, Inayoshi, Ingram,
  Inoue, Ioka, Isi, Isleif, Ito, Itoh, Iyer, Izumi, JaberianHamedan, Jacqmin,
  Jadhav, Jadhav, James, Jan, Jani, Janquart, Janssens, Janthalur, Jaranowski,
  Jariwala, Jaume, Jenkins, Jenner, Jeon, Jeunon, Jia, Jin, Johns,
  Johnson-McDaniel, Jones, Jones, Jones, Jones, Jones, Jonker, Ju, Jung, Jung,
  Junker, Juste, Kaihotsu, Kajita, Kakizaki, Kalaghatgi, Kalogera, Kamai,
  Kamiizumi, Kanda, Kandhasamy, Kang, Kanner, Kao, Kapadia, Kapasi, Karat,
  Karathanasis, Karki, Kashyap, Kasprzack, Kastaun, Katsanevas, Katsavounidis,
  Katzman, Kaur, Kawabe, Kawaguchi, Kawai, Kawasaki, Kéfélian, Keitel, Key,
  Khadka, Khalili, Khan, Khazanov, Khetan, Khursheed, Kijbunchoo, Kim, Kim,
  Kim, Kim, Kim, Kim, Kimball, Kimura, Kinley-Hanlon, Kirchhoff, Kissel, Kita,
  Kitazawa, Kleybolte, Klimenko, Knee, Knowles, Knyazev, Koch, Koekoek, Kojima,
  Kokeyama, Koley, Kolitsidou, Kolstein, Komori, Kondrashov, Kong, Kontos,
  Koper, Korobko, Kotake, Kovalam, Kozak, Kozakai, Kozu, Kringel, Krishnendu,
  Królak, Kuehn, Kuei, Kuijer, Kulkarni, Kumar, Kumar, Kumar, Kumar, Kume,
  Kuns, Kuo, Kuo, Kuromiya, Kuroyanagi, Kusayanagi, Kuwahara, Kwak, Lagabbe,
  Laghi, Lalande, Lam, Lamberts, Landry, Lane, Lang, Lange, Lantz, La~Rosa,
  Lartaux-Vollard, Lasky, Laxen, Lazzarini, Lazzaro, Leaci, Leavey, Lecoeuche,
  Lee, Lee, Lee, Lee, Lee, Lee, Lehmann, Lemaître, Leonardi, Leroy, Letendre,
  Levesque, Levin, Leviton, Leyde, Li, Li, Li, Li, Li, Li, Lin, Lin, Lin, Lin,
  Lin, Linde, Linker, Linley, Littenberg, Liu, Liu, Liu, Liu, Llamas,
  Llorens-Monteagudo, Lo, Lockwood, Loh, London, Longo, Lopez, Portilla,
  Lorenzini, Loriette, Lormand, Losurdo, Lott, Lough, Lousto, Lovelace,
  Lucaccioni, Lück, Lumaca, Lundgren, Luo, Lynam, Macas, MacInnis, Macleod,
  MacMillan, Macquet, Hernandez, Magazzù, Magee, Maggiore, Magnozzi, Mahesh,
  Majorana, Makarem, Maksimovic, Maliakal, Malik, Man, Mandic, Mangano, Mango,
  Mansell, Manske, Mantovani, Mapelli, Marchesoni, Marchio, Marion, Mark,
  Márka, Márka, Markakis, Markosyan, Markowitz, Maros, Marquina, Marsat,
  Martelli, Martin, Martin, Martinez, Martinez, Martinez, Martinovic, Martynov,
  Marx, Masalehdan, Mason, Massera, Masserot, Massinger, Masso-Reid,
  Mastrogiovanni, Matas, Mateu-Lucena, Matichard, Matiushechkina, Mavalvala,
  McCann, McCarthy, McClelland, McClincy, McCormick, McCuller, McGhee, McGuire,
  McIsaac, McIver, McRae, McWilliams, Meacher, Mehmet, Mehta, Meijer, Melatos,
  Melchor, Mendell, Menendez-Vazquez, Menoni, Mercer, Mereni, Merfeld, Merilh,
  Merritt, Merzougui, Meshkov, Messenger, Messick, Meyers, Meylahn, Mhaske,
  Miani, Miao, Michaloliakos, Michel, Michimura, Middleton, Milano, Miller,
  Miller, Miller, Millhouse, Mills, Milotti, Minazzoli, Minenkov, Mio, Mir,
  Miravet-Tenés, Mishra, Mishra, Mistry, Mitra, Mitrofanov, Mitselmakher,
  Mittleman, Miyakawa, Miyamoto, Miyazaki, Miyo, Miyoki, Mo, Modafferi, Moguel,
  Mogushi, Mohapatra, Mohite, Molina, Molina-Ruiz, Mondin, Montani, Moore,
  Moraru, Morawski, More, Moreno, Moreno, Mori, Morisaki, Moriwaki, Morrás,
  Mours, Mow-Lowry, Mozzon, Muciaccia, Mukherjee, Mukherjee, Mukherjee,
  Mukherjee, Mukherjee, Mukund, Mullavey, Munch, Muñiz, Murray, Musenich,
  Muusse, Nadji, Nagano, Nagano, Nagar, Nakamura, Nakano, Nakano, Nakashima,
  Nakayama, Napolano, Nardecchia, Narikawa, Naticchioni, Nayak, Nayak, Negishi,
  Neil, Neilson, Nelemans, Nelson, Nery, Neubauer, Neunzert, Ng, Ng, Nguyen,
  Nguyen, Nguyen, Quynh, Ni, Nichols, Nishizawa, Nissanke, Nitoglia, Nocera,
  Norman, North, Nozaki, Siles, Nuttall, Oberling, O’Brien, Obuchi, O’Dell,
  Oelker, Ogaki, Oganesyan, Oh, Oh, Oh, Ohashi, Ohishi, Ohkawa, Ohme, Ohta,
  Okada, Okutani, Okutomi, Olivetto, Oohara, Ooi, Oram, O’Reilly, Ormiston,
  Ormsby, Ortega, O’Shaughnessy, O’Shea, Oshino, Ossokine, Osthelder,
  Otabe, Ottaway, Overmier, Pace, Pagano, Page, Pagliaroli, Pai, Pai, Palamos,
  Palashov, Palomba, Pan, Pan, Panda, Pang, Pang, Pankow, Pannarale, Pant,
  Panther, Paoletti, Paoli, Paolone, Parisi, Park, Park, Parker, Pascucci,
  Pasqualetti, Passaquieti, Passuello, Patel, Pathak, Patricelli, Patron, Paul,
  Payne, Pedraza, Pegoraro, Pele, Arellano, Penn, Perego, Pereira, Pereira,
  Perez, Périgois, Perkins, Perreca, Perriès, Petermann, Petterson, Pfeiffer,
  Pham, Phukon, Piccinni, Pichot, Piendibene, Piergiovanni, Pierini, Pierro,
  Pillant, Pillas, Pilo, Pinard, Pinto, Pinto, Piotrzkowski, Piotrzkowski,
  Pirello, Pitkin, Placidi, Planas, Plastino, Pluchar, Poggiani, Polini, Pong,
  Ponrathnam, Popolizio, Porter, Poulton, Powell, Pracchia, Pradier, Prajapati,
  Prasai, Prasanna, Pratten, Principe, Prodi, Prokhorov, Prosposito, Prudenzi,
  Puecher, Punturo, Puosi, Puppo, Pürrer, Qi, Quetschke, Quitzow-James, Qutob,
  Raab, Raaijmakers, Radkins, Radulesco, Raffai, Rail, Raja, Rajan, Ramirez,
  Ramirez, Ramos-Buades, Rana, Rapagnani, Rapol, Ray, Raymond, Raza, Razzano,
  Read, Rees, Regimbau, Rei, Reid, Reid, Reitze, Relton, Renzini, Rettegno,
  Reza, Rezac, Ricci, Richards, Richardson, Richardson, Riemenschneider, Riles,
  Rinaldi, Rink, Rizzo, Robertson, Robie, Robinet, Rocchi, Rodriguez, Rolland,
  Rollins, Romanelli, Romano, Romel, Romero-Rodríguez, Romero-Shaw, Romie,
  Ronchini, Rosa, Rose, Rosińska, Ross, Rowan, Rowlinson, Roy, Roy, Roy,
  Rozza, Ruggi, Ruiz-Rocha, Ryan, Sachdev, Sadecki, Sadiq, Sago, Saito, Saito,
  Sakai, Sakai, Sakellariadou, Sakuno, Salafia, Salconi, Saleem, Salemi,
  Samajdar, Sanchez, Sanchez, Sanchez, Sanchis-Gual, Sanders, Sanuy, Saravanan,
  Sarin, Sassolas, Satari, Sathyaprakash, Sato, Sato, Sauter, Savage, Sawada,
  Sawant, Sawant, Sayah, Schaetzl, Scheel, Scheuer, Schiworski, Schmidt,
  Schmidt, Schnabel, Schneewind, Schofield, Schönbeck, Schulte, Schutz,
  Schwartz, Scott, Scott, Seglar-Arroyo, Sekiguchi, Sekiguchi, Sellers,
  Sengupta, Sentenac, Seo, Sequino, Sergeev, Setyawati, Shaffer, Shahriar,
  Shams, Shao, Sharma, Sharma, Shawhan, Shcheblanov, Shibagaki, Shikauchi,
  Shimizu, Shimoda, Shimode, Shinkai, Shishido, Shoda, Shoemaker, Shoemaker,
  ShyamSundar, Sieniawska, Sigg, Singer, Singh, Singh, Singha, Sintes, Sipala,
  Skliris, Slagmolen, Slaven-Blair, Smetana, Smith, Smith, Soldateschi, Somala,
  Somiya, Son, Soni, Soni, Sordini, Sorrentino, Sorrentino, Sotani, Soulard,
  Souradeep, Sowell, Spagnuolo, Spencer, Spera, Srinivasan, Srivastava,
  Srivastava, Staats, Stachie, Steer, Steinhoff, Steinlechner, Steinlechner,
  Stevenson, Stops, Stover, Strain, Strang, Stratta, Strunk, Sturani, Stuver,
  Sudhagar, Sudhir, Sugimoto, Suh, Sullivan, Sullivan, Summerscales, Sun, Sun,
  Sunil, Sur, Suresh, Sutton, Suzuki, Suzuki, Swinkels, Szczepańczyk,
  Szewczyk, Tacca, Tagoshi, Tait, Takahashi, Takahashi, Takamori, Takano,
  Takeda, Takeda, Talbot, Talbot, Tanaka, Tanaka, Tanaka, Tanaka, Tanaka,
  Tanasijczuk, Tanioka, Tanner, Tao, Tao, Martín, Taranto, Tasson, Telada,
  Tenorio, Terhune, Terkowski, Thirugnanasambandam, Thomas, Thomas, Thomas,
  Thompson, Thondapu, Thorne, Thrane, Tiwari, Tiwari, Tiwari, Toivonen, Toland,
  Tolley, Tomaru, Tomigami, Tomura, Tonelli, Torres-Forné, Torrie, e~Melo,
  Töyrä, Trapananti, Travasso, Traylor, Trevor, Tringali, Tripathee, Troiano,
  Trovato, Trozzo, Trudeau, Tsai, Tsai, Tsang, Tsang, Tsao, Tse, Tso, Tsubono,
  Tsuchida, Tsukada, Tsuna, Tsutsui, Tsuzuki, Turbang, Turconi, Tuyenbayev,
  Ubhi, Uchikata, Uchiyama, Udall, Ueda, Uehara, Ueno, Ueshima, Unnikrishnan,
  Uraguchi, Urban, Ushiba, Utina, Vahlbruch, Vajente, Vajpeyi, Valdes,
  Valentini, Valsan, van Bakel, van Beuzekom, van~den Brand, Van Den~Broeck,
  Vander-Hyde, van~der Schaaf, van Heijningen, Vanosky, van Putten, van
  Remortel, Vardaro, Vargas, Varma, Vasúth, Vecchio, Vedovato, Veitch, Veitch,
  Venneberg, Venugopalan, Verkindt, Verma, Verma, Veske, Vetrano, Viceré,
  Vidyant, Viets, Vijaykumar, Villa-Ortega, Vinet, Virtuoso, Vitale, Vo, Vocca,
  von Reis, von Wrangel, Vorvick, Vyatchanin, Wade, Wade, Wagner, Walet,
  Walker, Wallace, Wallace, Walsh, Wang, Wang, Wang, Ward, Warner, Was,
  Washimi, Washington, Watchi, Weaver, Webster, Weinert, Weinstein, Weiss,
  Weller, Weller, Wellmann, Wen, Weßels, Wette, Whelan, White, Whiting,
  Whittle, Wilken, Williams, Williams, Williams, Williamson, Willis, Willke,
  Wilson, Winkler, Wipf, Wlodarczyk, Woan, Woehler, Wofford, Wong, Wu, Wu, Wu,
  Wu, Wysocki, Xiao, Xu, Yamada, Yamamoto, Yamamoto, Yamamoto, Yamamoto,
  Yamashita, Yamazaki, Yang, Yang, Yang, Yang, Yang, Yap, Yeeles, Yelikar,
  Ying, Yokogawa, Yokoyama, Yokozawa, Yoo, Yoshioka, Yu, Yu, Yuzurihara,
  Zadrożny, Zanolin, Zeidler, Zelenova, Zendri, Zevin, Zhan, Zhang, Zhang,
  Zhang, Zhang, Zhang, Zhao, Zhao, Zhao, Zhao, Zheng, Zhou, Zhou, Zhu, Zhu,
  Zimmerman, Zlochower, Zucker, \& Zweizig}]{gwtc-3}
Abbott, R., Abbott, T., Acernese, F., {et~al.} 2023{\natexlab{a}}, Physical
  Review X, 13, \dodoi{10.1103/physrevx.13.041039}

\bibitem[{Abbott {et~al.}(2023{\natexlab{b}})Abbott, Abe, Acernese, Ackley,
  Adhikari, Adhikari, Adkins, Adya, Affeldt, Agarwal, Agathos, Agatsuma,
  Aggarwal, Aguiar, Aiello, Ain, Ajith, Akutsu, Albanesi, Alfaidi, Allocca,
  Altin, Amato, Anand, Anand, Ananyeva, Anderson, Anderson, Ando, Andrade,
  Andres, Andrés-Carcasona, Andrić, Angelova, Ansoldi, Antelis, Antier,
  Apostolatos, Appavuravther, Appert, Apple, Arai, Araya, Araya, Areeda,
  Arène, Aritomi, Arnaud, Arogeti, Aronson, Arun, Asada, Asali, Ashton, Aso,
  Assiduo, de~Souza~Melo, Aston, Astone, Aubin, AultONeal, Austin, Babak,
  Badaracco, Bader, Badger, Bae, Bae, Baer, Bagnasco, Bai, Baird, Bajpai, Baka,
  Ball, Ballardin, Ballmer, Balsamo, Baltus, Banagiri, Banerjee, Bankar,
  Barayoga, Barbieri, Barbieri, Barish, Barker, Barneo, Barone, Barr, Barsotti,
  Barsuglia, Barta, Bartlett, Barton, Bartos, Basak, Bassiri, Basti, Bawaj,
  Bayley, Bazzan, Becher, Bécsy, Bedakihale, Beirnaert, Bejger, Belahcene,
  Benedetto, Beniwal, Benjamin, Bennett, Bentley, BenYaala, Bera, Berbel,
  Bergamin, Berger, Bernuzzi, Berry, Bersanetti, Bertolini, Betzwieser,
  Beveridge, Bhandare, Bhandari, Bhardwaj, Bhatt, Bhattacharjee, Bhaumik,
  Bianchi, Bilenko, Billingsley, Bilicki, Bini, Birney, Birnholtz, Biscans,
  Bischi, Biscoveanu, Bisht, Biswas, Bitossi, Bizouard, Blackburn, Blair,
  Blair, Blair, Bobba, Bode, Boër, Bogaert, Boldrini, Bolingbroke, Bonavena,
  Bondu, Bonilla, Bonnand, Booker, Boom, Bork, Boschi, Bose, Bose, Bossilkov,
  Boudart, Bouffanais, Bozzi, Bradaschia, Brady, Bramley, Branch, Branchesi,
  Brau, Breschi, Briant, Briggs, Brillet, Brinkmann, Brockill, Brooks, Brooks,
  Brown, Brunett, Bruno, Bruntz, Bryant, Bucci, Bulik, Bulten, Buonanno,
  Burtnyk, Buscicchio, Buskulic, Buy, Byer, Davies, Cabras, Cabrita, Cadonati,
  Caesar, Cagnoli, Cahillane, Bustillo, Callaghan, Callister, Calloni, Cameron,
  Camp, Canepa, Canevarolo, Cannavacciuolo, Cannon, Cao, Cao, Capocasa, Capote,
  Carapella, Carbognani, Carlassara, Carlin, Carney, Carpinelli, Carrillo,
  Carullo, Carver, Diaz, Casentini, Castaldi, Caudill, Cavaglià, Cavalier,
  Cavalieri, Cella, Cerdá-Durán, Cesarini, Chaibi, Subrahmanya, Champion,
  Chan, Chan, Chan, Chan, Chan, Chandra, Chang, Chanial, Chao, Chapman-Bird,
  Charlton, Chase, Chassande-Mottin, Chatterjee, Chatterjee, Chatterjee,
  Chaturvedi, Chaty, Chatziioannou, Chen, Chen, Chen, Chen, Chen, Chen, Chen,
  Chen, Chen, Cheng, Cheong, Cheung, Chia, Chiadini, Chiang, Chiarini,
  Chierici, Chincarini, Chiofalo, Chiummo, Choudhary, Choudhary, Christensen,
  Chu, Chu, Chua, Chung, Ciani, Ciecielag, Cieślar, Cifaldi, Ciobanu, Ciolfi,
  Cipriano, Clara, Clark, Clearwater, Clesse, Cleva, Coccia, Codazzo, Cohadon,
  Cohen, Colleoni, Collette, Colombo, Colpi, Compton, Constancio, Conti,
  Cooper, Corban, Corbitt, Cordero-Carrión, Corezzi, Corley, Cornish, Corre,
  Corsi, Cortese, Costa, Cotesta, Cottingham, Coughlin, Coulon, Countryman,
  Cousins, Couvares, Coward, Cowart, Coyne, Coyne, Creighton, Creighton,
  Criswell, Croquette, Crowder, Cudell, Cullen, Cumming, Cummings, Cunningham,
  Cuoco, Curyło, Dabadie, Dal~Canton, Dall’Osso, Dálya, Dana, D’Angelo,
  Danilishin, D’Antonio, Danzmann, Darsow-Fromm, Dasgupta, Datrier, Datta,
  Datta, Dattilo, Dave, Davier, Davis, Davis, Daw, De~Alarc’on, Dean, DeBra,
  Deenadayalan, Degallaix, De~Laurentis, Deléglise, Del~Favero, De~Lillo,
  De~Lillo, Dell’Aquila, Del~Pozzo, DeMarchi, De~Matteis, D’Emilio, Demos,
  Dent, Depasse, De~Pietri, De~Rosa, De~Rossi, DeSalvo, De~Simone, Dhurandhar,
  Díaz, Didio, Dietrich, Fiore, Fronzo, Giorgio, Giovanni, Giovanni, Girolamo,
  Lieto, Michele, Ding, Pace, Palma, Renzo, Divakarla, Dmitriev, Doctor,
  Donahue, D’Onofrio, Donovan, Dooley, Doravari, Drago, Driggers, Drori,
  Ducoin, Dupej, Dupletsa, Durante, D’Urso, Duverne, Dwyer, Eassa, Easter,
  Ebersold, Eckhardt, Eddolls, Edelman, Edo, Edy, Effler, Eguchi, Eichholz,
  Eikenberry, Eisenmann, Eisenstein, Ejlli, Engelby, Enomoto, Errico, Essick,
  Estellés, Estevez, Etienne, Etzel, Evans, Evans, Evstafyeva, Ewing, Fabrizi,
  Faedi, Fafone, Fair, Fairhurst, Fan, Farah, Farinon, Farr, Farr,
  Fauchon-Jones, Favaro, Favata, Fays, Fazio, Feicht, Fejer, Fenyvesi,
  Ferguson, Fernandez-Galiana, Ferrante, Ferreira, Fidecaro, Figura, Fiori,
  Fiori, Fishbach, Fisher, Fittipaldi, Fiumara, Flaminio, Floden, Fong, Font,
  Fornal, Forsyth, Franke, Frasca, Frasconi, Freed, Frei, Freise, Freitas,
  Frey, Fritschel, Frolov, Fronzé, Fujii, Fujikawa, Fujimoto, Fulda, Fyffe,
  Gabbard, Gadre, Gair, Gais, Galaudage, Gamba, Ganapathy, Ganguly, Gao,
  Gaonkar, Garaventa, García~Núñez, García-Quirós, Garufi, Gateley,
  Gayathri, Ge, Gemme, Gennai, George, Gerberding, Gergely, Gewecke, Ghonge,
  Ghosh, Ghosh, Ghosh, Ghosh, Ghosh, Giacomazzo, Giacoppo, Giaime, Giardina,
  Gibson, Gier, Giesler, Giri, Gissi, Gkaitatzis, Glanzer, Gleckl, Godwin,
  Goetz, Goetz, Gohlke, Golomb, Goncharov, González, Gosselin, Gouaty, Gould,
  Goyal, Grace, Grado, Graham, Granata, Granata, Grant, Gras, Grassia, Gray,
  Gray, Greco, Green, Green, Gretarsson, Gretarsson, Griffith, Griffiths,
  Griggs, Grignani, Grimaldi, Grimes, Grimm, Grote, Grunewald, Gruning, Gruson,
  Guerra, Guidi, Guimaraes, Guixé, Gulati, Gunny, Guo, Guo, Gupta, Gupta,
  Gupta, Gupta, Gupta, Gustafson, Guzman, Ha, Hadiputrawan, Haegel, Haino,
  Halim, Hall, Hamilton, Hammond, Han, Haney, Hanks, Hanna, Hannam, Hannuksela,
  Hansen, Hansen, Hanson, Harder, Haris, Harms, Harry, Harry, Hartwig,
  Hasegawa, Haskell, Haster, Hathaway, Hattori, Haughian, Hayakawa, Hayama,
  Hayes, Healy, Heidmann, Heidt, Heintze, Heinze, Heinzel, Heitmann, Hellman,
  Hello, Helmling-Cornell, Hemming, Hendry, Heng, Hennes, Hennig, Hennig,
  Henshaw, Hernandez, Vivanco, Heurs, Hewitt, Higginbotham, Hild, Hill,
  Himemoto, Hines, Hirata, Hirose, Ho, Hochheim, Hofman, Hohmann, Holcomb,
  Holland, Hollows, Holmes, Holt, Holz, Hong, Hough, Hourihane, Howell, Hoy,
  Hoyland, Hreibi, Hsieh, Hsieh, Hsiung, Hsu, Huang, Huang, Huang, Huang,
  Huang, Huang, Hübner, Huddart, Hughey, Hui, Hui, Husa, Huttner, Huxford,
  Huynh-Dinh, Ide, Idzkowski, Iess, Inayoshi, Inoue, Iosif, Isi, Isleif, Ito,
  Itoh, Iyer, JaberianHamedan, Jacqmin, Jacquet, Jadhav, Jadhav, Jain, James,
  Jan, Jani, Janquart, Janssens, Janthalur, Jaranowski, Jariwala, Jaume,
  Jenkins, Jenner, Jeon, Jia, Jiang, Jin, Johns, Johnston, Jones, Jones, Jones,
  Jones, Joshi, Ju, Jue, Jung, Jung, Junker, Juste, Kaihotsu, Kajita, Kakizaki,
  Kalaghatgi, Kalogera, Kamai, Kamiizumi, Kanda, Kandhasamy, Kang, Kanner, Kao,
  Kapadia, Kapasi, Karathanasis, Karki, Kashyap, Kasprzack, Kastaun, Kato,
  Katsanevas, Katsavounidis, Katzman, Kaur, Kawabe, Kawaguchi, Kéfélian,
  Keitel, Key, Khadka, Khalili, Khan, Khanam, Khazanov, Khetan, Khursheed,
  Kijbunchoo, Kim, Kim, Kim, Kim, Kim, Kim, Kim, Kimball, Kimura,
  Kinley-Hanlon, Kirchhoff, Kissel, Klimenko, Klinger, Knee, Knowles, Knust,
  Knyazev, Kobayashi, Koch, Koekoek, Kohri, Kokeyama, Koley, Kolitsidou,
  Kolstein, Komori, Kondrashov, Kong, Kontos, Koper, Korobko, Kovalam, Koyama,
  Kozak, Kozakai, Kringel, Królak, Kuehn, Kuei, Kuijer, Kulkarni, Kumar,
  Kumar, Kumar, Kumar, Kume, Kuns, Kuromiya, Kuroyanagi, Kwak, Lacaille,
  Lagabbe, Laghi, Lalande, Lalleman, Lam, Lamberts, Landry, Lane, Lang, Lange,
  Lantz, Rosa, Lartaux-Vollard, Lasky, Laxen, Lazzarini, Lazzaro, Leaci,
  Leavey, LeBohec, Lecoeuche, Lee, Lee, Lee, Lee, Lee, Legred, Lehmann,
  Lemaître, Lenti, Leonardi, Leonova, Leroy, Letendre, Levesque, Levin,
  Leviton, Leyde, Li, Li, Li, Li, Li, Li, Li, Lin, Lin, Lin, Lin, Lin, Lin,
  Linde, Linker, Linley, Littenberg, Liu, Liu, Liu, Liu, Llamas, Lo, Lo,
  London, Longo, Lopez, Portilla, Lorenzini, Loriette, Lormand, Losurdo, Lott,
  Lough, Lousto, Lovelace, Lucaccioni, Lück, Lumaca, Lundgren, Luo, Lynam,
  Ma’arif, Macas, Machtinger, MacInnis, Macleod, MacMillan, Macquet,
  Hernandez, Magazzù, Magee, Maggiore, Magnozzi, Mahesh, Majorana, Maksimovic,
  Maliakal, Malik, Man, Mandic, Mangano, Mansell, Manske, Mantovani, Mapelli,
  Marchesoni, Marín~Pina, Marion, Mark, Márka, Márka, Markakis, Markosyan,
  Markowitz, Maros, Marquina, Marsat, Martelli, Martin, Martin, Martinez,
  Martinez, Martinez, Martinovic, Martynov, Marx, Masalehdan, Mason, Massera,
  Masserot, Masso-Reid, Mastrogiovanni, Matas, Mateu-Lucena, Matichard,
  Matiushechkina, Mavalvala, McCann, McCarthy, McClelland, McClincy, McCormick,
  McCuller, McGhee, McGuire, McIsaac, McIver, McRae, McWilliams, Meacher,
  Mehmet, Mehta, Meijer, Melatos, Melchor, Mendell, Menendez-Vazquez, Menoni,
  Mercer, Mereni, Merfeld, Merilh, Merritt, Merzougui, Meshkov, Messenger,
  Messick, Meyers, Meylahn, Mhaske, Miani, Miao, Michaloliakos, Michel,
  Michimura, Middleton, Mihaylov, Milano, Miller, Miller, Miller, Millhouse,
  Mills, Milotti, Minenkov, Mio, Mir, Miravet-Tenés, Mishkin, Mishra, Mishra,
  Mistry, Mitra, Mitrofanov, Mitselmakher, Mittleman, Miyakawa, Miyo, Miyoki,
  Mo, Modafferi, Moguel, Mogushi, Mohapatra, Mohite, Molina, Molina-Ruiz,
  Mondin, Montani, Moore, Moragues, Moraru, Morawski, More, More, Moreno,
  Moreno, Mori, Morisaki, Morisue, Moriwaki, Mours, Mow-Lowry, Mozzon,
  Muciaccia, Mukherjee, Mukherjee, Mukherjee, Mukherjee, Mukherjee, Mukund,
  Mullavey, Munch, Muñiz, Murray, Musenich, Muusse, Nadji, Nagano, Nagar,
  Nakamura, Nakano, Nakano, Nakayama, Napolano, Nardecchia, Narikawa, Narola,
  Naticchioni, Nayak, Nayak, Neil, Neilson, Nelson, Nelson, Nery, Neubauer,
  Neunzert, Ng, Ng, Nguyen, Nguyen, Nguyen, Quynh, Ni, Ni, Nichols, Nishimoto,
  Nishizawa, Nissanke, Nitoglia, Nocera, Norman, North, Nozaki, Nurbek,
  Nuttall, Obayashi, Oberling, O’Brien, O’Dell, Oelker, Ogaki, Oganesyan,
  Oh, Oh, Oh, Ohashi, Ohashi, Ohkawa, Ohme, Ohta, Okada, Okutani, Olivetto,
  Oohara, Oram, O’Reilly, Ormiston, Ormsby, O’Shaughnessy, O’Shea,
  Oshino, Ossokine, Osthelder, Otabe, Ottaway, Overmier, Pace, Pagano, Pagano,
  Page, Pagliaroli, Pai, Pai, Pal, Palamos, Palashov, Palomba, Pan, Pan, Panda,
  Pang, Pankow, Pannarale, Pant, Panther, Paoletti, Paoli, Paolone, Pappas,
  Parisi, Park, Park, Parker, Pascucci, Pasqualetti, Passaquieti, Passuello,
  Patel, Pathak, Patricelli, Patron, Paul, Payne, Pedraza, Pedurand, Pegoraro,
  Pele, Arellano, Penano, Penn, Perego, Pereira, Pereira, Perez, Périgois,
  Perkins, Perreca, Perriès, Pesios, Petermann, Petterson, Pfeiffer, Pham,
  Pham, Phukon, Phurailatpam, Piccinni, Pichot, Piendibene, Piergiovanni,
  Pierini, Pierro, Pillant, Pillas, Pilo, Pinard, Pineda-Bosque, Pinto, Pinto,
  Piotrzkowski, Piotrzkowski, Pirello, Pitkin, Placidi, Placidi, Planas,
  Plastino, Pluchar, Poggiani, Polini, Pong, Ponrathnam, Porter, Poulton,
  Poverman, Powell, Pracchia, Pradier, Prajapati, Prasai, Prasanna, Pratten,
  Principe, Prodi, Prokhorov, Prosposito, Prudenzi, Puecher, Punturo, Puosi,
  Puppo, Pürrer, Qi, Quartey, Quetschke, Quinonez, Quitzow-James, Raab,
  Raaijmakers, Radkins, Radulesco, Raffai, Rail, Raja, Rajan, Ramirez, Ramirez,
  Ramos-Buades, Rana, Rapagnani, Ray, Raymond, Raza, Razzano, Read, Rees,
  Regimbau, Rei, Reid, Reid, Reitze, Relton, Renzini, Rettegno, Revenu, Reza,
  Rezac, Ricci, Richards, Richardson, Richardson, Riemenschneider, Riles,
  Rinaldi, Rink, Robertson, Robie, Robinet, Rocchi, Rodriguez, Rolland,
  Rollins, Romanelli, Romano, Romel, Romero, Romero-Shaw, Romie, Ronchini,
  Rosa, Rose, Rosińska, Ross, Rowan, Rowlinson, Roy, Roy, Roy, Rozza, Ruggi,
  Ruiz-Rocha, Ryan, Sachdev, Sadecki, Sadiq, Saha, Saito, Sakai, Sakellariadou,
  Sakon, Salafia, Salces-Carcoba, Salconi, Saleem, Salemi, Samajdar, Sanchez,
  Sanchez, Sanchez, Sanchis-Gual, Sanders, Sanuy, Saravanan, Sarin, Sassolas,
  Satari, Sathyaprakash, Sauter, Savage, Savant, Sawada, Sawant, Sayah,
  Schaetzl, Scheel, Scheuer, Schiworski, Schmidt, Schmidt, Schnabel,
  Schneewind, Schofield, Schönbeck, Schulte, Schutz, Schwartz, Scott, Scott,
  Seglar-Arroyo, Sekiguchi, Sellers, Sengupta, Sentenac, Seo, Sequino, Sergeev,
  Setyawati, Shaffer, Shahriar, Shaikh, Shams, Shao, Sharma, Sharma, Shawhan,
  Shcheblanov, Sheela, Shikano, Shikauchi, Shimizu, Shimode, Shinkai, Shishido,
  Shoda, Shoemaker, Shoemaker, ShyamSundar, Sieniawska, Sigg, Silenzi, Singer,
  Singh, Singh, Singh, Singha, Sintes, Sipala, Skliris, Slagmolen,
  Slaven-Blair, Smetana, Smith, Smith, Smith, Soldateschi, Somala, Somiya,
  Song, Soni, Sordini, Sorrentino, Sorrentino, Soulard, Souradeep, Sowell,
  Spagnuolo, Spencer, Spera, Spinicelli, Srivastava, Srivastava, Staats,
  Stachie, Stachurski, Steer, Steinlechner, Steinlechner, Stergioulas, Stops,
  Stover, Strain, Strang, Stratta, Strong, Strunk, Sturani, Stuver, Suchenek,
  Sudhagar, Sudhir, Sugimoto, Suh, Sullivan, Summerscales, Sun, Sunil, Sur,
  Suresh, Sutton, Suzuki, Suzuki, Suzuki, Swinkels, Szczepańczyk, Szewczyk,
  Tacca, Tagoshi, Tait, Takahashi, Takahashi, Takano, Takeda, Takeda, Talbot,
  Talbot, Tamanini, Tanaka, Tanaka, Tanaka, Tanasijczuk, Tanioka, Tanner, Tao,
  Tao, Tapia, Tapia San~Martín, Taranto, Taruya, Tasson, Tenorio, Terhune,
  Terkowski, Thirugnanasambandam, Thomas, Thomas, Thompson, Thompson, Thondapu,
  Thorne, Thrane, Tiwari, Tiwari, Tiwari, Toivonen, Tolley, Tomaru, Tomura,
  Tonelli, Tornasi, Torres-Forné, Torrie, e~Melo, Töyrä, Trapananti,
  Travasso, Traylor, Trevor, Tringali, Tripathee, Troiano, Trovato, Trozzo,
  Trudeau, Tsai, Tsang, Tsang, Tsao, Tse, Tso, Tsuchida, Tsukada, Tsuna,
  Tsutsui, Turbang, Turconi, Turski, Tuyenbayev, Ubhi, Uchikata, Uchiyama,
  Udall, Ueda, Uehara, Ueno, Ueshima, Unnikrishnan, Urban, Ushiba, Utina,
  Vajente, Vajpeyi, Valdes, Valentini, Valsan, Bakel, Beuzekom, Dael, van~den
  Brand, Van Den~Broeck, Vander-Hyde, van Haevermaet, van Heijningen, Putten,
  Remortel, Vardaro, Vargas, Varma, Vasúth, Vecchio, Vedovato, Veitch, Veitch,
  Venneberg, Venugopalan, Verkindt, Verma, Verma, Vermeulen, Veske, Vetrano,
  Viceré, Vidyant, Viets, Vijaykumar, Villa-Ortega, Vinet, Virtuoso, Vitale,
  Vocca, Reis, Wrangel, Vorvick, Vyatchanin, Wade, Wade, Wagner, Walet, Walker,
  Wallace, Wallace, Wang, Wang, Wang, Ward, Warner, Was, Washimi, Washington,
  Watchi, Weaver, Weaving, Webster, Weinert, Weinstein, Weiss, Weller, Weller,
  Wellmann, Wen, Weßels, Wette, Whelan, White, Whiting, Whittle, Wilken,
  Williams, Williams, Williamson, Willis, Willke, Wilson, Wipf, Wlodarczyk,
  Woan, Woehler, Wofford, Wong, Wong, Wright, Wu, Wu, Wu, Wysocki, Xiao,
  Yamada, Yamamoto, Yamamoto, Yamamoto, Yamashita, Yamazaki, Yang, Yang, Yang,
  Yang, Yang, Yang, Yap, Yeeles, Yeh, Yelikar, Ying, Yokoyama, Yokozawa, Yoo,
  Yoshioka, Yu, Yu, Yuzurihara, Zadrożny, Zanolin, Zeidler, Zelenova, Zendri,
  Zevin, Zhan, Zhang, Zhang, Zhang, Zhang, Zhang, Zhang, Zhao, Zhao, Zhao,
  Zhao, Zhou, Zhou, Zhu, Zhu, Zimmerman, Zucker, \& Zweizig}]{o3_cosmology}
Abbott, R., Abe, H., Acernese, F., {et~al.} 2023{\natexlab{b}}, The
  Astrophysical Journal, 949, 76, \dodoi{10.3847/1538-4357/ac74bb}

\bibitem[{Abbott {et~al.}(2023{\natexlab{c}})Abbott, Abbott, Acernese, Ackley,
  Adams, Adhikari, Adhikari, Adya, Affeldt, Agarwal, Agathos, Agatsuma,
  Aggarwal, Aguiar, Aiello, Ain, Ajith, Akutsu, de~Alarc\'on, Akcay, Albanesi,
  Allocca, Altin, Amato, Anand, Anand, Ananyeva, Anderson, Anderson, Ando,
  Andrade, Andres, Andri\ifmmode~\acute{c}\else \'{c}\fi{}, Angelova, Ansoldi,
  Antelis, Antier, Antonini, Appert, Arai, Arai, Arai, Araki, Araya, Araya,
  Areeda, Ar\`ene, Aritomi, Arnaud, Arogeti, Aronson, Arun, Asada, Asali,
  Ashton, Aso, Assiduo, Aston, Astone, Aubin, Austin, Babak, Badaracco, Bader,
  Badger, Bae, Bae, Baer, Bagnasco, Bai, Baiotti, Baird, Bajpai, Ball,
  Ballardin, Ballmer, Balsamo, Baltus, Banagiri, Bankar, Barayoga, Barbieri,
  Barish, Barker, Barneo, Barone, Barr, Barsotti, Barsuglia, Barta, Bartlett,
  Barton, Bartos, Bassiri, Basti, Bawaj, Bayley, Baylor, Bazzan, B\'ecsy,
  Bedakihale, Bejger, Belahcene, Benedetto, Beniwal, Bennett, Bentley,
  BenYaala, Bergamin, Berger, Bernuzzi, Berry, Bersanetti, Bertolini,
  Betzwieser, Beveridge, Bhandare, Bhardwaj, Bhattacharjee, Bhaumik, Bilenko,
  Billingsley, Bini, Birney, Birnholtz, Biscans, Bischi, Biscoveanu, Bisht,
  Biswas, Bitossi, Bizouard, Blackburn, Blair, Blair, Blair, Bobba, Bode, Boer,
  Bogaert, Boldrini, Bonavena, Bondu, Bonilla, Bonnand, Booker, Boom, Bork,
  Boschi, Bose, Bose, Bossilkov, Boudart, Bouffanais, Bozzi, Bradaschia, Brady,
  Bramley, Branch, Branchesi, Brandt, Brau, Breschi, Briant, Briggs, Brillet,
  Brinkmann, Brockill, Brooks, Brooks, Brown, Brunett, Bruno, Bruntz, Bryant,
  Bulik, Bulten, Buonanno, Buscicchio, Buskulic, Buy, Byer, Cadonati, Cagnoli,
  Cahillane, Bustillo, Callaghan, Callister, Calloni, Cameron, Camp, Canepa,
  Canevarolo, Cannavacciuolo, Cannon, Cao, Cao, Capocasa, Capote, Carapella,
  Carbognani, Carlin, Carney, Carpinelli, Carrillo, Carullo, Carver, Diaz,
  Casentini, Castaldi, Caudill, Cavagli\`a, Cavalier, Cavalieri, Ceasar, Cella,
  Cerd\'a-Dur\'an, Cesarini, Chaibi, Chakravarti, Subrahmanya, Champion, Chan,
  Chan, Chan, Chan, Chan, Chandra, Chanial, Chao, Chapman-Bird, Charlton,
  Chase, Chassande-Mottin, Chatterjee, Chatterjee, Chatterjee, Chaturvedi,
  Chaty, Chatziioannou, Chen, Chen, Chen, Chen, Chen, Chen, Chen, Chen, Cheng,
  Cheong, Cheung, Chia, Chiadini, Chiang, Chiarini, Chierici, Chincarini,
  Chiofalo, Chiummo, Cho, Cho, Choudhary, Choudhary, Christensen, Chu, Chu,
  Chu, Chua, Chung, Ciani, Ciecielag, Cie\ifmmode~\acute{s}\else \'{s}\fi{}lar,
  Cifaldi, Ciobanu, Ciolfi, Cipriano, Cirone, Clara, Clark, Clark, Clarke,
  Clearwater, Clesse, Cleva, Coccia, Codazzo, Cohadon, Cohen, Cohen, Colleoni,
  Collette, Colombo, Colpi, Compton, Constancio, Conti, Cooper, Corban,
  Corbitt, Cordero-Carri\'on, Corezzi, Corley, Cornish, Corre, Corsi, Cortese,
  Costa, Cotesta, Coughlin, Coulon, Countryman, Cousins, Couvares, Coward,
  Cowart, Coyne, Coyne, Creighton, Creighton, Criswell, Croquette, Crowder,
  Cudell, Cullen, Cumming, Cummings, Cunningham, Cuoco, Cury\l{}o, Dabadie,
  Canton, Dall'Osso, D\'alya, Dana, DaneshgaranBajastani, D'Angelo, Danila,
  Danilishin, D'Antonio, Danzmann, Darsow-Fromm, Dasgupta, Datrier, Datta,
  Dattilo, Dave, Davier, Davies, Davis, Davis, Daw, Dean, DeBra, Deenadayalan,
  Degallaix, De~Laurentis, Del\'eglise, Del~Favero, De~Lillo, De~Lillo,
  Del~Pozzo, DeMarchi, De~Matteis, D'Emilio, Demos, Dent, Depasse, De~Pietri,
  De~Rosa, De~Rossi, DeSalvo, De~Simone, Dhurandhar, D\'{\i}az, Diaz-Ortiz,
  Didio, Dietrich, Di~Fiore, Di~Fronzo, Di~Giorgio, Di~Giovanni, Di~Giovanni,
  Di~Girolamo, Di~Lieto, Ding, Di~Pace, Di~Palma, Di~Renzo, Divakarla,
  Dmitriev, Doctor, D'Onofrio, Donovan, Dooley, Doravari, Dorrington, Drago,
  Driggers, Drori, Ducoin, Dupej, Durante, D'Urso, Duverne, Dwyer, Eassa,
  Easter, Ebersold, Eckhardt, Eddolls, Edelman, Edo, Edy, Effler, Eguchi,
  Eichholz, Eikenberry, Eisenmann, Eisenstein, Ejlli, Engelby, Enomoto, Errico,
  Essick, Estell\'es, Estevez, Etienne, Etzel, Evans, Evans, Ewing, Fafone,
  Fair, Fairhurst, Farah, Farinon, Farr, Farr, Farrow, Fauchon-Jones, Favaro,
  Favata, Fays, Fazio, Feicht, Fejer, Fenyvesi, Ferguson, Fernandez-Galiana,
  Ferrante, Ferreira, Fidecaro, Figura, Fiori, Fishbach, Fisher, Fittipaldi,
  Fiumara, Flaminio, Floden, Fong, Font, Fornal, Forsyth, Franke, Frasca,
  Frasconi, Frederick, Freed, Frei, Freise, Frey, Fritschel, Frolov, Fronz\'e,
  Fujii, Fujikawa, Fukunaga, Fukushima, Fulda, Fyffe, Gabbard, Gadre, Gair,
  Gais, Galaudage, Gamba, Ganapathy, Ganguly, Gao, Gaonkar, Garaventa,
  Garc\'{\i}a, Garc\'{\i}a-N\'u\~nez, Garc\'{\i}a-Quir\'os, Garufi, Gateley,
  Gaudio, Gayathri, Ge, Gemme, Gennai, George, George, Gerberding, Gergely,
  Gewecke, Ghonge, Ghosh, Ghosh, Ghosh, Ghosh, Giacomazzo, Giacoppo, Giaime,
  Giardina, Gibson, Gier, Giesler, Giri, Gissi, Glanzer, Gleckl, Godwin,
  Golomb, Goetz, Goetz, Gohlke, Goncharov, Gonz\'alez, Gopakumar, Gosselin,
  Gouaty, Gould, Grace, Grado, Granata, Granata, Grant, Gras, Grassia, Gray,
  Gray, Greco, Green, Green, Gretarsson, Gretarsson, Griffith, Griffiths,
  Griggs, Grignani, Grimaldi, Grimm, Grote, Grunewald, Gruning, Guerra, Guidi,
  Guimaraes, Guix\'e, Gulati, Guo, Guo, Gupta, Gupta, Gupta, Gustafson,
  Gustafson, Guzman, Ha, Haegel, Hagiwara, Haino, Halim, Hall, Hamilton,
  Hammond, Han, Haney, Hanks, Hanna, Hannam, Hannuksela, Hansen, Hansen,
  Hanson, Harder, Hardwick, Haris, Harms, Harry, Harry, Hartwig, Hasegawa,
  Haskell, Hasskew, Haster, Hattori, Haughian, Hayakawa, Hayama, Hayes, Healy,
  Heidmann, Heidt, Heintze, Heinze, Heinzel, Heitmann, Hellman, Hello,
  Helmling-Cornell, Hemming, Hendry, Heng, Hennes, Hennig, Hennig, Hernandez,
  Vivanco, Heurs, Hild, Hill, Himemoto, Hines, Hiranuma, Hirata, Hirose,
  Hochheim, Hofman, Hohmann, Holcomb, Holland, Hollows, Holmes, Holt, Holz,
  Hong, Hopkins, Hough, Hourihane, Howell, Hoy, Hoyland, Hreibi, Hsieh, Hsu,
  Huang, Huang, Huang, Huang, Huang, Huang, H\"ubner, Huddart, Hughey, Hui,
  Hui, Husa, Huttner, Huxford, Huynh-Dinh, Ide, Idzkowski, Iess, Ikenoue, Imam,
  Inayoshi, Ingram, Inoue, Ioka, Isi, Isleif, Ito, Itoh, Iyer, Izumi,
  JaberianHamedan, Jacqmin, Jadhav, Jadhav, James, Jan, Jani, Janquart,
  Janssens, Janthalur, Jaranowski, Jariwala, Jaume, Jenkins, Jenner, Jeon,
  Jeunon, Jia, Jin, Johns, Jones, Jones, Jones, Jones, Jones, Jonker, Ju, Jung,
  Jung, Junker, Juste, Kaihotsu, Kajita, Kakizaki, Kalaghatgi, Kalogera, Kamai,
  Kamiizumi, Kanda, Kandhasamy, Kang, Kanner, Kao, Kapadia, Kapasi, Karat,
  Karathanasis, Karki, Kashyap, Kasprzack, Kastaun, Katsanevas, Katsavounidis,
  Katzman, Kaur, Kawabe, Kawaguchi, Kawai, Kawasaki, K\'ef\'elian, Keitel, Key,
  Khadka, Khalili, Khan, Khazanov, Khetan, Khursheed, Kijbunchoo, Kim, Kim,
  Kim, Kim, Kim, Kim, Kimball, Kimura, Kinley-Hanlon, Kirchhoff, Kissel, Kita,
  Kitazawa, Kleybolte, Klimenko, Knee, Knowles, Knyazev, Koch, Koekoek, Kojima,
  Kokeyama, Koley, Kolitsidou, Kolstein, Komori, Kondrashov, Kong, Kontos,
  Koper, Korobko, Kotake, Kovalam, Kozak, Kozakai, Kozu, Kringel, Krishnendu,
  Kr\'olak, Kuehn, Kuei, Kuijer, Kulkarni, Kumar, Kumar, Kumar, Kumar, Kume,
  Kuns, Kuo, Kuo, Kuromiya, Kuroyanagi, Kusayanagi, Kuwahara, Kwak, Lagabbe,
  Laghi, Lalande, Lam, Lamberts, Landry, Landry, Lane, Lang, Lange, Lantz,
  La~Rosa, Lartaux-Vollard, Lasky, Laxen, Lazzarini, Lazzaro, Leaci, Leavey,
  Lecoeuche, Lee, Lee, Lee, Lee, Lee, Lee, Lehmann, Lema\^{\i}tre, Leonardi,
  Leroy, Letendre, Levesque, Levin, Leviton, Leyde, Li, Li, Li, Li, Li, Li,
  Lin, Lin, Lin, Lin, Lin, Linde, Linker, Linley, Littenberg, Liu, Liu, Liu,
  Liu, Llamas, Llorens-Monteagudo, Lo, Lockwood, Loh, London, Longo, Lopez,
  Portilla, Lorenzini, Loriette, Lormand, Losurdo, Lott, Lough, Lousto,
  Lovelace, Lucaccioni, L\"uck, Lumaca, Lundgren, Luo, Lynam, Macas, MacInnis,
  Macleod, MacMillan, Macquet, Hernandez, Magazz\`u, Magee, Maggiore, Magnozzi,
  Mahesh, Majorana, Makarem, Maksimovic, Maliakal, Malik, Man, Mandic, Mangano,
  Mango, Mansell, Manske, Mantovani, Mapelli, Marchesoni, Marchio, Marion,
  Mark, M\'arka, M\'arka, Markakis, Markosyan, Markowitz, Maros, Marquina,
  Marsat, Martelli, Martin, Martin, Martinez, Martinez, Martinez, Martinovic,
  Martynov, Marx, Masalehdan, Mason, Massera, Masserot, Massinger, Masso-Reid,
  Mastrogiovanni, Matas, Mateu-Lucena, Matichard, Matiushechkina, Mavalvala,
  McCann, McCarthy, McClelland, McClincy, McCormick, McCuller, McGhee, McGuire,
  McIsaac, McIver, McRae, McWilliams, Meacher, Mehmet, Mehta, Meijer, Melatos,
  Melchor, Mendell, Menendez-Vazquez, Menoni, Mercer, Mereni, Merfeld, Merilh,
  Merritt, Merzougui, Meshkov, Messenger, Messick, Meyers, Meylahn, Mhaske,
  Miani, Miao, Michaloliakos, Michel, Michimura, Middleton, Milano, Miller,
  Miller, Miller, Miller, Millhouse, Mills, Milotti, Minazzoli, Minenkov, Mio,
  Mir, Miravet-Ten\'es, Mishra, Mishra, Mistry, Mitra, Mitrofanov,
  Mitselmakher, Mittleman, Miyakawa, Miyamoto, Miyazaki, Miyo, Miyoki, Mo,
  Modafferi, Moguel, Mogushi, Mohapatra, Mohite, Molina, Molina-Ruiz, Mondin,
  Montani, Moore, Moraru, Morawski, More, Moreno, Moreno, Mori, Morisaki,
  Moriwaki, Morr\'as, Mours, Mow-Lowry, Mozzon, Muciaccia, Mukherjee,
  Mukherjee, Mukherjee, Mukherjee, Mukherjee, Mukund, Mullavey, Munch, Mu\~niz,
  Murray, Musenich, Muusse, Nadji, Nagano, Nagano, Nagar, Nakamura, Nakano,
  Nakano, Nakashima, Nakayama, Napolano, Nardecchia, Narikawa, Naticchioni,
  Nayak, Nayak, Negishi, Neil, Neilson, Nelemans, Nelson, Nery, Neubauer,
  Neunzert, Ng, Ng, Nguyen, Nguyen, Nguyen, Quynh, Ni, Nichols, Nishizawa,
  Nissanke, Nitoglia, Nocera, Norman, North, Nozaki, Siles, Nuttall, Oberling,
  O'Brien, Obuchi, O'Dell, Oelker, Ogaki, Oganesyan, Oh, Oh, Oh, Ohashi,
  Ohishi, Ohkawa, Ohme, Ohta, Okada, Okutani, Okutomi, Olivetto, Oohara, Ooi,
  Oram, O'Reilly, Ormiston, Ormsby, Ortega, O'Shaughnessy, O'Shea, Oshino,
  Ossokine, Osthelder, Otabe, Ottaway, Overmier, Pace, Pagano, Page,
  Pagliaroli, Pai, Pai, Palamos, Palashov, Palomba, Pan, Pan, Panda, Pang,
  Pang, Pankow, Pannarale, Pant, Panther, Paoletti, Paoli, Paolone, Parisi,
  Park, Park, Parker, Pascucci, Pasqualetti, Passaquieti, Passuello, Patel,
  Pathak, Patricelli, Patron, Paul, Payne, Pedraza, Pegoraro, Pele, Arellano,
  Penn, Perego, Pereira, Pereira, Perez, P\'erigois, Perkins, Perreca,
  Perri\`es, Petermann, Petterson, Pfeiffer, Pham, Phukon, Piccinni, Pichot,
  Piendibene, Piergiovanni, Pierini, Pierro, Pillant, Pillas, Pilo, Pinard,
  Pinto, Pinto, Piotrzkowski, Piotrzkowski, Pirello, Pitkin, Placidi, Planas,
  Plastino, Pluchar, Poggiani, Polini, Pong, Ponrathnam, Popolizio, Porter,
  Poulton, Powell, Pracchia, Pradier, Prajapati, Prasai, Prasanna, Pratten,
  Principe, Prodi, Prokhorov, Prosposito, Prudenzi, Puecher, Punturo, Puosi,
  Puppo, P\"urrer, Qi, Quetschke, Quitzow-James, Raab, Raaijmakers, Radkins,
  Radulesco, Raffai, Rail, Raja, Rajan, Ramirez, Ramirez, Ramos-Buades, Rana,
  Rapagnani, Rapol, Ray, Raymond, Raza, Razzano, Read, Rees, Regimbau, Rei,
  Reid, Reid, Reitze, Relton, Renzini, Rettegno, Reza, Rezac, Ricci, Richards,
  Richardson, Richardson, Riemenschneider, Riles, Rinaldi, Rink, Rizzo,
  Robertson, Robie, Robinet, Rocchi, Rodriguez, Rolland, Rollins, Romanelli,
  Romano, Romel, Romero-Rodr\'{\i}guez, Romero-Shaw, Romie, Ronchini, Rosa,
  Rose, Rosi\ifmmode~\acute{n}\else \'{n}\fi{}ska, Ross, Rowan, Rowlinson, Roy,
  Roy, Roy, Rozza, Ruggi, Ryan, Sachdev, Sadecki, Sadiq, Sago, Saito, Saito,
  Sakai, Sakai, Sakellariadou, Sakuno, Salafia, Salconi, Saleem, Salemi,
  Samajdar, Sanchez, Sanchez, Sanchez, Sanchis-Gual, Sanders, Sanuy, Saravanan,
  Sarin, Sassolas, Satari, Sathyaprakash, Sato, Sato, Sauter, Savage, Sawada,
  Sawant, Sawant, Sayah, Schaetzl, Scheel, Scheuer, Schiworski, Schmidt,
  Schmidt, Schnabel, Schneewind, Schofield, Sch\"onbeck, Schulte, Schutz,
  Schwartz, Scott, Scott, Seglar-Arroyo, Sekiguchi, Sekiguchi, Sellers,
  Sengupta, Sentenac, Seo, Sequino, Sergeev, Setyawati, Shaffer, Shahriar,
  Shams, Shao, Sharma, Sharma, Shawhan, Shcheblanov, Shibagaki, Shikauchi,
  Shimizu, Shimoda, Shimode, Shinkai, Shishido, Shoda, Shoemaker, Shoemaker,
  ShyamSundar, Sieniawska, Sigg, Singer, Singh, Singh, Singha, Sintes, Sipala,
  Skliris, Slagmolen, Slaven-Blair, Smetana, Smith, Smith, Soldateschi, Somala,
  Somiya, Son, Soni, Soni, Sordini, Sorrentino, Sorrentino, Sotani, Soulard,
  Souradeep, Sowell, Spagnuolo, Spencer, Spera, Srinivasan, Srivastava,
  Srivastava, Staats, Stachie, Steer, Steinhoff, Steinlechner, Steinlechner,
  Stevenson, Stops, Stover, Strain, Strang, Stratta, Strunk, Sturani, Stuver,
  Sudhagar, Sudhir, Sugimoto, Suh, Sullivan, Summerscales, Sun, Sun, Sunil,
  Sur, Suresh, Sutton, Suzuki, Suzuki, Swinkels, Szczepa\ifmmode~\acute{n}\else
  \'{n}\fi{}czyk, Szewczyk, Tacca, Tagoshi, Tait, Takahashi, Takahashi,
  Takamori, Takano, Takeda, Takeda, Talbot, Talbot, Tanaka, Tanaka, Tanaka,
  Tanaka, Tanaka, Tanasijczuk, Tanioka, Tanner, Tao, Tao, Mart\'{\i}n, Taranto,
  Tasson, Telada, Tenorio, Terhune, Terkowski, Thirugnanasambandam, Thomas,
  Thomas, Thomas, Thompson, Thondapu, Thorne, Thrane, Tiwari, Tiwari, Tiwari,
  Toivonen, Toland, Tolley, Tomaru, Tomigami, Tomura, Tonelli, Torres-Forn\'e,
  Torrie, e~Melo, T\"oyr\"a, Trapananti, Travasso, Traylor, Trevor, Tringali,
  Tripathee, Troiano, Trovato, Trozzo, Trudeau, Tsai, Tsai, Tsang, Tsang, Tsao,
  Tse, Tso, Tsubono, Tsuchida, Tsukada, Tsuna, Tsutsui, Tsuzuki, Turbang,
  Turconi, Tuyenbayev, Ubhi, Uchikata, Uchiyama, Udall, Ueda, Uehara, Ueno,
  Ueshima, Unnikrishnan, Uraguchi, Urban, Ushiba, Utina, Vahlbruch, Vajente,
  Vajpeyi, Valdes, Valentini, Valsan, van Bakel, van Beuzekom, van~den Brand,
  Van Den~Broeck, Vander-Hyde, van~der Schaaf, van Heijningen, Vanosky, van
  Putten, van Remortel, Vardaro, Vargas, Varma, Vas\'uth, Vecchio, Vedovato,
  Veitch, Veitch, Venneberg, Venugopalan, Verkindt, Verma, Verma, Veske,
  Vetrano, Vicer\'e, Vidyant, Viets, Vijaykumar, Villa-Ortega, Vinet, Virtuoso,
  Vitale, Vo, Vocca, von Reis, von Wrangel, Vorvick, Vyatchanin, Wade, Wade,
  Wagner, Walet, Walker, Wallace, Wallace, Walsh, Wang, Wang, Wang, Ward,
  Warner, Was, Washimi, Washington, Watchi, Weaver, Webster, Weinert,
  Weinstein, Weiss, Weller, Wellmann, Wen, We\ss{}els, Wette, Whelan, White,
  Whiting, Whittle, Wilken, Williams, Williams, Williamson, Willis, Willke,
  Wilson, Winkler, Wipf, Wlodarczyk, Woan, Woehler, Wofford, Wong, Wu, Wu, Wu,
  Wu, Wysocki, Xiao, Xu, Yamada, Yamamoto, Yamamoto, Yamamoto, Yamamoto,
  Yamashita, Yamazaki, Yang, Yang, Yang, Yang, Yang, Yap, Yeeles, Yelikar,
  Ying, Yokogawa, Yokoyama, Yokozawa, Yoo, Yoshioka, Yu, Yu, Yuzurihara,
  Zadro\ifmmode~\dot{z}\else \.{z}\fi{}ny, Zanolin, Zeidler, Zelenova, Zendri,
  Zevin, Zhan, Zhang, Zhang, Zhang, Zhang, Zhang, Zhao, Zhao, Zhao, Zhao,
  Zheng, Zhou, Zhou, Zhu, Zhu, Zimmerman, Zlochower, Zucker, \&
  Zweizig}]{o3_randp}
Abbott, R., Abbott, T.~D., Acernese, F., {et~al.} 2023{\natexlab{c}}, Phys.
  Rev. X, 13, 011048, \dodoi{10.1103/PhysRevX.13.011048}

\bibitem[{Abbott {et~al.}(2024)}]{LIGOScientific:2023bwz}
Abbott, R., {et~al.} 2024, Astrophys. J., 970, 191,
  \dodoi{10.3847/1538-4357/ad3e83}

\bibitem[{Acernese {et~al.}(2015)}]{virgo_detector}
Acernese, F., {et~al.} 2015, Class. Quant. Grav., 32, 024001,
  \dodoi{10.1088/0264-9381/32/2/024001}

\bibitem[{Akutsu {et~al.}(2021)}]{kagra_detector}
Akutsu, T., {et~al.} 2021, PTEP, 2021, 05A101, \dodoi{10.1093/ptep/ptaa125}

\bibitem[{Allen(2005)}]{Allen:2004gu}
Allen, B. 2005, Phys. Rev. D, 71, 062001, \dodoi{10.1103/PhysRevD.71.062001}

\bibitem[{Allen {et~al.}(2012)Allen, Anderson, Brady, Brown, \&
  Creighton}]{Allen:2005fk}
Allen, B., Anderson, W.~G., Brady, P.~R., Brown, D.~A., \& Creighton, J. D.~E.
  2012, Phys. Rev. D, 85, 122006, \dodoi{10.1103/PhysRevD.85.122006}

\bibitem[{Ashton {et~al.}(2019)}]{bilby_paper}
Ashton, G., {et~al.} 2019, Astrophys. J. Suppl., 241, 27,
  \dodoi{10.3847/1538-4365/ab06fc}

\bibitem[{Ashton {et~al.}(2024)}]{cbcflow}
---. 2024, {\textsc{CBCFlow}}.
\newblock \url{https://cbc.docs.ligo.org/projects/cbcflow/index.html}

\bibitem[{Barsode {et~al.}(2025)Barsode, Goyal, \& Ajith}]{Barsode:2024zwv}
Barsode, A., Goyal, S., \& Ajith, P. 2025, Astrophys. J., 980, 258,
  \dodoi{10.3847/1538-4357/adae10}

\bibitem[{Binney \& Tremaine(1989)}]{Binney:1989je}
Binney, J., \& Tremaine, S. 1989, Galactic dynamics,
  \dodoi{10.1515/9781400828722}

\bibitem[{Biwer {et~al.}(2019)Biwer, Capano, De, Cabero, Brown, Nitz, \&
  Raymond}]{Biwer:2018osg}
Biwer, C.~M., Capano, C.~D., De, S., {et~al.} 2019, Publ. Astron. Soc. Pac.,
  131, 024503, \dodoi{10.1088/1538-3873/aaef0b}

\bibitem[{Cannon {et~al.}(2021)Cannon, Caudill, Chan, Cousins, Creighton,
  Ewing, Fong, Godwin, Hanna, Hooper, Huxford, Magee, Meacher, Messick,
  Morisaki, Mukherjee, Ohta, Pace, Privitera, {de Ruiter}, Sachdev, Singer,
  Singh, Tapia, Tsukada, Tsuna, Tsutsui, Ueno, Viets, Wade, \&
  Wade}]{CANNON2021100680}
Cannon, K., Caudill, S., Chan, C., {et~al.} 2021, SoftwareX, 14, 100680,
  \dodoi{https://doi.org/10.1016/j.softx.2021.100680}

\bibitem[{Cao {et~al.}(2014)Cao, Li, \& Wang}]{Cao:2014oaa}
Cao, Z., Li, L.-F., \& Wang, Y. 2014, Phys. Rev. D, 90, 062003,
  \dodoi{10.1103/PhysRevD.90.062003}

\bibitem[{Chakraborty \& Mukherjee(2025)}]{Chakraborty:2025maj}
Chakraborty, A., \& Mukherjee, S. 2025.
\newblock \doarXiv{2503.16281}

\bibitem[{Cheung {et~al.}(2021)Cheung, Gais, Hannuksela, \&
  Li}]{Cheung:2020okf}
Cheung, M. H.~Y., Gais, J., Hannuksela, O.~A., \& Li, T. G.~F. 2021, Mon. Not.
  Roy. Astron. Soc., 503, 3326, \dodoi{10.1093/mnras/stab579}

\bibitem[{Christian {et~al.}(2018)Christian, Vitale, \&
  Loeb}]{Christian:2018vsi}
Christian, P., Vitale, S., \& Loeb, A. 2018, Phys. Rev. D, 98, 103022,
  \dodoi{10.1103/PhysRevD.98.103022}

\bibitem[{Collaboration(2022)}]{o4_psd}
Collaboration, L.-V.-K. 2022, Noise Curves for use in simulations pre-O4, Tech.
  rep.
\newblock \url{https://dcc.ligo.org/LIGO-T2200043/public}

\bibitem[{Dai {et~al.}(2018)Dai, Li, Zackay, Mao, \& Lu}]{Dai:2018enj}
Dai, L., Li, S.-S., Zackay, B., Mao, S., \& Lu, Y. 2018, Phys. Rev. D, 98,
  104029, \dodoi{10.1103/PhysRevD.98.104029}

\bibitem[{Dai {et~al.}(2017)Dai, Venumadhav, \& Sigurdson}]{Dai:2016igl}
Dai, L., Venumadhav, T., \& Sigurdson, K. 2017, Phys. Rev. D, 95, 044011,
  \dodoi{10.1103/PhysRevD.95.044011}

\bibitem[{Dal~Canton {et~al.}(2014)}]{DalCanton:2014hxh}
Dal~Canton, T., {et~al.} 2014, Phys. Rev. D, 90, 082004,
  \dodoi{10.1103/PhysRevD.90.082004}

\bibitem[{Deguchi \& Watson(1986)}]{Deguchi:1986zz}
Deguchi, S., \& Watson, W.~D. 1986, Phys. Rev. D, 34, 1708,
  \dodoi{10.1103/PhysRevD.34.1708}

\bibitem[{Diego {et~al.}(2019)Diego, Hannuksela, Kelly, Broadhurst, Kim, Li,
  Smoot, \& Pagano}]{Diego:2019lcd}
Diego, J., Hannuksela, O., Kelly, P., {et~al.} 2019, Astron. Astrophys., 627,
  A130, \dodoi{10.1051/0004-6361/201935490}

\bibitem[{Diego(2020)}]{Diego:2019rzc}
Diego, J.~M. 2020, Phys. Rev. D, 101, 123512,
  \dodoi{10.1103/PhysRevD.101.123512}

\bibitem[{Ezquiaga {et~al.}(2021)Ezquiaga, Holz, Hu, Lagos, \&
  Wald}]{Ezquiaga:2020gdt}
Ezquiaga, J.~M., Holz, D.~E., Hu, W., Lagos, M., \& Wald, R.~M. 2021, Phys.
  Rev. D, 103, 6, \dodoi{10.1103/PhysRevD.103.064047}

\bibitem[{Ezquiaga {et~al.}(2023)Ezquiaga, Hu, \& Lo}]{phazap}
Ezquiaga, J.~M., Hu, W., \& Lo, R. K.~L. 2023, Physical Review D, 108,
  \dodoi{10.1103/physrevd.108.103520}

\bibitem[{Goyal {et~al.}(2021)Goyal, D., Kapadia, \& Ajith}]{goyal_lensid}
Goyal, S., D., H., Kapadia, S.~J., \& Ajith, P. 2021, Physical Review D, 104,
  \dodoi{10.1103/physrevd.104.124057}

\bibitem[{Goyal {et~al.}(2024)Goyal, Kapadia, Cudell, Li, \&
  Chan}]{Goyal:2023lqf}
Goyal, S., Kapadia, S.~J., Cudell, J.-R., Li, A. K.~Y., \& Chan, J. C.~L. 2024,
  Phys. Rev. D, 109, 023028, \dodoi{10.1103/PhysRevD.109.023028}

\bibitem[{Hannuksela {et~al.}(2019)Hannuksela, Haris, Ng, Kumar, Mehta, Keitel,
  Li, \& Ajith}]{Hannuksela:2019kle}
Hannuksela, O.~A., Haris, K., Ng, K. K.~Y., {et~al.} 2019, Astrophys. J. Lett.,
  874, L2, \dodoi{10.3847/2041-8213/ab0c0f}

\bibitem[{Haris {et~al.}(2018)Haris, Mehta, Kumar, Venumadhav, \&
  Ajith}]{haris_posterior_overlap}
Haris, K., Mehta, A.~K., Kumar, S., Venumadhav, T., \& Ajith, P. 2018,
  Identifying strongly lensed gravitational wave signals from binary black hole
  mergers.
\newblock \doarXiv{1807.07062}

\bibitem[{Janquart {et~al.}(2021{\natexlab{a}})Janquart, Hannuksela, K., \& Van
  Den~Broeck}]{Janquart:2021qov}
Janquart, J., Hannuksela, O.~A., K., H., \& Van Den~Broeck, C.
  2021{\natexlab{a}}, Mon. Not. Roy. Astron. Soc.,
  \dodoi{10.1093/mnras/stab1991}

\bibitem[{Janquart {et~al.}(2023{\natexlab{a}})Janquart, Haris, Hannuksela, \&
  Van Den~Broeck}]{Janquart:2023osz}
Janquart, J., Haris, K., Hannuksela, O.~A., \& Van Den~Broeck, C.
  2023{\natexlab{a}}, Mon. Not. Roy. Astron. Soc., 526, 3088,
  \dodoi{10.1093/mnras/stad2838}

\bibitem[{Janquart {et~al.}(2021{\natexlab{b}})Janquart, Seo, Hannuksela, Li,
  \& Broeck}]{Janquart:2021nus}
Janquart, J., Seo, E., Hannuksela, O.~A., Li, T. G.~F., \& Broeck, C. V.~D.
  2021{\natexlab{b}}, Astrophys. J. Lett., 923, L1,
  \dodoi{10.3847/2041-8213/ac3bcf}

\bibitem[{Janquart {et~al.}(2023{\natexlab{b}})Janquart, Wright, Goyal,
  {et~al.}}]{Janquart:2023mvf}
Janquart, J., Wright, M., Goyal, S., {et~al.} 2023{\natexlab{b}}, Mon. Not.
  Roy. Astron. Soc., 526, 3832, \dodoi{10.1093/mnras/stad2909}

\bibitem[{Janquart {et~al.}(2024)}]{Janquart:2024ztv}
Janquart, J., {et~al.} 2024, \dodoi{10.1093/mnras/staf049}

\bibitem[{Jung \& Shin(2019)}]{Jung:2017flg}
Jung, S., \& Shin, C.~S. 2019, Phys. Rev. Lett., 122, 041103,
  \dodoi{10.1103/PhysRevLett.122.041103}

\bibitem[{Lai {et~al.}(2018)Lai, Hannuksela, Herrera-Mart{\'\i}n, Diego,
  Broadhurst, \& Li}]{Lai:2018rto}
Lai, K.-H., Hannuksela, O.~A., Herrera-Mart{\'\i}n, A., {et~al.} 2018, Phys.
  Rev. D, 98, 083005, \dodoi{10.1103/PhysRevD.98.083005}

\bibitem[{Li {et~al.}(2023{\natexlab{a}})Li, Chan, Fong, Chong, Weinstein, \&
  Ezquiaga}]{Li:2023zdl}
Li, A. K.~Y., Chan, J. C.~L., Fong, H., {et~al.} 2023{\natexlab{a}}.
\newblock \doarXiv{2311.06416}

\bibitem[{Li {et~al.}(2023{\natexlab{b}})Li, Lo, Sachdev, Chan, Lin, Li, \&
  Weinstein}]{Li:2019osa}
Li, A. K.~Y., Lo, R. K.~L., Sachdev, S., {et~al.} 2023{\natexlab{b}}, Phys.
  Rev. D, 107, 123014, \dodoi{10.1103/PhysRevD.107.123014}

\bibitem[{Litzkow {et~al.}(1988)Litzkow, Livny, \& Mutka}]{condor-hunter}
Litzkow, M., Livny, M., \& Mutka, M. 1988, in Proceedings of the 8th
  International Conference of Distributed Computing Systems

\bibitem[{Liu {et~al.}(2023)Liu, Wong, Leong, More, Hannuksela, \&
  Li}]{Liu:2023ikc}
Liu, A., Wong, I. C.~F., Leong, S. H.~W., {et~al.} 2023, Mon. Not. Roy. Astron.
  Soc., 525, 4149, \dodoi{10.1093/mnras/stad1302}

\bibitem[{Lo \& Magana~Hernandez(2023)}]{Lo:2021nae}
Lo, R. K.~L., \& Magana~Hernandez, I. 2023, Phys. Rev. D, 107, 123015,
  \dodoi{10.1103/PhysRevD.107.123015}

\bibitem[{{LVK Collaborations}(2025)}]{gracedb_detections}
{LVK Collaborations}. 2025, GraceDB.
\newblock \url{https://gracedb.ligo.org/superevents/public/O4/}

\bibitem[{McIsaac {et~al.}(2020)McIsaac, Keitel, Collett, Harry, Mozzon, Edy,
  \& Bacon}]{McIsaac:2019use}
McIsaac, C., Keitel, D., Collett, T., {et~al.} 2020, Phys. Rev. D, 102, 084031,
  \dodoi{10.1103/PhysRevD.102.084031}

\bibitem[{Meena {et~al.}(2022)Meena, Mishra, More, Bose, \&
  Bagla}]{Meena:2022unp}
Meena, A.~K., Mishra, A., More, A., Bose, S., \& Bagla, J.~S. 2022, Mon. Not.
  Roy. Astron. Soc., 517, 872, \dodoi{10.1093/mnras/stac2721}

\bibitem[{Messick {et~al.}(2017)}]{Messick:2016aqy}
Messick, C., {et~al.} 2017, Phys. Rev. D, 95, 042001,
  \dodoi{10.1103/PhysRevD.95.042001}

\bibitem[{{Mishra} {et~al.}(2021){Mishra}, {Meena}, {More}, {Bose}, \&
  {Bagla}}]{mishra2021}
{Mishra}, A., {Meena}, A.~K., {More}, A., {Bose}, S., \& {Bagla}, J.~S. 2021,
  \mnras, 508, 4869, \dodoi{10.1093/mnras/stab2875}

\bibitem[{More \& More(2022)}]{more_statistics}
More, A., \& More, S. 2022, Monthly Notices of the Royal Astronomical Society,
  515, 1044–1051, \dodoi{10.1093/mnras/stac1704}

\bibitem[{Nakamura(1998)}]{Nakamura:1997sw}
Nakamura, T.~T. 1998, Phys. Rev. Lett., 80, 1138,
  \dodoi{10.1103/PhysRevLett.80.1138}

\bibitem[{Nakamura \& Deguchi(1999)}]{Nakamura:1999uwi}
Nakamura, T.~T., \& Deguchi, S. 1999, Prog. Theor. Phys. Suppl., 133, 137,
  \dodoi{10.1143/ptps.133.137}

\bibitem[{Nitz {et~al.}(2017)Nitz, Dent, Dal~Canton, Fairhurst, \&
  Brown}]{Nitz:2017svb}
Nitz, A.~H., Dent, T., Dal~Canton, T., Fairhurst, S., \& Brown, D.~A. 2017,
  Astrophys. J., 849, 118, \dodoi{10.3847/1538-4357/aa8f50}

\bibitem[{Ohanian(1974)}]{Ohanian:1974ys}
Ohanian, H. 1974, Int. J. Theor. Phys., 9, 425, \dodoi{10.1007/BF01810927}

\bibitem[{Pagano {et~al.}(2020)Pagano, Hannuksela, \& Li}]{Pagano:2020rwj}
Pagano, G., Hannuksela, O.~A., \& Li, T. G.~F. 2020, Astron. Astrophys., 643,
  A167, \dodoi{10.1051/0004-6361/202038730}

\bibitem[{Pratten {et~al.}(2021)}]{Pratten:2020ceb}
Pratten, G., {et~al.} 2021, Phys. Rev. D, 103, 104056,
  \dodoi{10.1103/PhysRevD.103.104056}

\bibitem[{Romero-Shaw {et~al.}(2020)}]{bilby_pipe_paper}
Romero-Shaw, I.~M., {et~al.} 2020, Mon. Not. Roy. Astron. Soc., 499, 3295,
  \dodoi{10.1093/mnras/staa2850}

\bibitem[{Sachdev {et~al.}(2019)}]{Sachdev:2019vvd}
Sachdev, S., {et~al.} 2019.
\newblock \doarXiv{1901.08580}

\bibitem[{Schneider {et~al.}(1992)Schneider, Ehlers, \&
  Falco}]{Schneider:1992bmb}
Schneider, P., Ehlers, J., \& Falco, E.~E. 1992, {Gravitational Lenses},
  Astronomy and Astrophysics Library (Springer),
  \dodoi{10.1007/978-3-662-03758-4}

\bibitem[{{Shapiro}(1964)}]{shapiro_delay}
{Shapiro}, I.~I. 1964, \prl, 13, 789, \dodoi{10.1103/PhysRevLett.13.789}

\bibitem[{Singer \& Price(2016)}]{Singer:2015ema}
Singer, L.~P., \& Price, L.~R. 2016, Phys. Rev. D, 93, 024013,
  \dodoi{10.1103/PhysRevD.93.024013}

\bibitem[{Smith {et~al.}(2020)Smith, Ashton, Vajpeyi, \& Talbot}]{pbilby_paper}
Smith, R. J.~E., Ashton, G., Vajpeyi, A., \& Talbot, C. 2020, Mon. Not. Roy.
  Astron. Soc., 498, 4492, \dodoi{10.1093/mnras/staa2483}

\bibitem[{Takahashi \& Nakamura(2003)}]{Takahashi:2003ix}
Takahashi, R., \& Nakamura, T. 2003, Astrophys. J., 595, 1039,
  \dodoi{10.1086/377430}

\bibitem[{Thorne(1982)}]{Thorne:1982cv}
Thorne, K. 1982, in {Les Houches Summer School on Gravitational Radiation},
  1--57

\bibitem[{Tsukada {et~al.}(2023)}]{Tsukada:2023edh}
Tsukada, L., {et~al.} 2023, Phys. Rev. D, 108, 043004,
  \dodoi{10.1103/PhysRevD.108.043004}

\bibitem[{Vijaykumar {et~al.}(2022)Vijaykumar, Mehta, \&
  Ganguly}]{Vijaykumar:2022dlp}
Vijaykumar, A., Mehta, A.~K., \& Ganguly, A. 2022.
\newblock \doarXiv{2202.06334}

\bibitem[{Wang {et~al.}(2021)Wang, Lo, Li, \& Chen}]{Wang:2021kzt}
Wang, Y., Lo, R. K.~L., Li, A. K.~Y., \& Chen, Y. 2021, Phys. Rev. D, 103,
  104055, \dodoi{10.1103/PhysRevD.103.104055}

\bibitem[{Williams(2025)}]{asimov_docs}
Williams, D. 2025, \textsc{Asimov} documentation.
\newblock \url{https://asimov.docs.ligo.org/asimov/master/index.html}

\bibitem[{Williams {et~al.}(2023)Williams, Veitch, Chiofalo, Schmidt, Udall,
  Vajpeji, \& Hoy}]{asimov}
Williams, D., Veitch, J., Chiofalo, M.~L., {et~al.} 2023, Journal of Open
  Source Software, 8, 4170, \dodoi{10.21105/joss.04170}

\bibitem[{Wright \& Hendry(2022)}]{Wright:2021cbn}
Wright, M., \& Hendry, M. 2022, The Astrophysical Journal, 935, 68,
  \dodoi{10.3847/1538-4357/ac7ec2}

\bibitem[{Wright {et~al.}(2025)Wright, Janquart, \& Cremonese}]{lensingflow}
Wright, M., Janquart, J., \& Cremonese, P. 2025, \textsc{LensingFlow}.
\newblock \url{https://git.ligo.org/lensingflow/lensingflow}

\bibitem[{Wright {et~al.}(2021)Wright, Liu, Seo, Wong, \&
  Janquart}]{gravelamps}
Wright, M., Liu, A., Seo, E., Wong, I. C.~F., \& Janquart, J. 2021,
  \textsc{Gravelamps}.
\newblock \url{https://git.ligo.org/mick.wright/Gravelamps}

\bibitem[{Yeung {et~al.}(2021)Yeung, Cheung, Gais, Hannuksela, \&
  Li}]{Yeung:2021roe}
Yeung, S. M.~C., Cheung, M. H.~Y., Gais, J. A.~J., Hannuksela, O.~A., \& Li, T.
  G.~F. 2021.
\newblock \doarXiv{2112.07635}

\end{thebibliography}

\end{document}